# Fast covariance-free spatiotemporal modeling via coarse-to-fine learning

Daisuke Murakami

Department of Fundamental Statistical Mathematics, Institute of Statistical Mathematics, Japan

(E-mail: dmuraka@ism.ac.jp)

Scalable spatiotemporal modeling remains challenging because conventional methods rely on covariance models that tightly couple spatial representation and temporal inference, often leading to high computational costs. To address this difficulty, we developed coarse-to-fine spatiotemporal modeling (CF-STM), a framework that extends coarse-to-fine spatial modeling (CF-SM) to spatiotemporal settings. CF-STM represents latent spatial processes through multiscale locally weighted models, with temporal dependence modeled separately through state-space models defined at local centers. This covariance-free formulation achieves scalable computation while allowing temporal models to be flexibly specified without altering the spatial representation. The spatiotemporal covariance implied by this construction is characterized as well, and reduces to a separable spatiotemporal Matérn covariance in an idealized limit. Monte Carlo experiments demonstrate predictive performance comparable to that of alternative scalable spatiotemporal models at a substantially lower computational cost. An application to long-term residential land price data in the Tokyo metropolitan area shows that CF-STM flexibly captures complex spatiotemporal patterns while enabling interpretable inferences. CF-STM is implemented in an R package spCF.

## 1. Introduction

Geographic phenomena are increasingly observed at fine spatial and temporal resolutions across environmental and socioeconomic domains. These data make it possible to model dynamic spatiotemporal processes and their associated uncertainties. However, many statistical models that incorporate spatiotemporal dependence tend to be computationally demanding and difficult to scale to large datasets.

A common approach to modeling spatiotemporal processes is to jointly represent spatial and temporal dependence through spatial Gaussian process models combined with state-space models that describe temporal dynamics (Cressie and Wikle 2011). However, the computational cost of Gaussian processes scales poorly for large datasets because inference relies on repeated operations on large covariance matrices. Scalability has been substantially improved by approximating spatial Gaussian processes, including basis function approximations (e.g., Cressie and Johannesson, 2008; Katzfuss, 2017; Tzeng and Huang, 2018), sparse covariance/precision approximations, including nearest-neighbor Gaussian processes (NNGP; Datta et al., 2016a), sparse generalized Vecchia (Katzfuss and Guinness, 2021), and stochastic partial differential equation (SPDE) methods (Lindgren et al., 2011).

These spatial models have been extended and combined in various ways to jointly capture dependence over space and time. A widely used approach embeds random spatial effects within a hierarchical dynamic state-space formulation, where temporal evolution is described through state equations (Cressie and Wikle, 2011; Wikle et al., 2019). The scalable spatial approximations noted above have been applied to space-time modeling, including dynamic models with multiscale spatial basis functions (Cressie et al., 2010; Jurek and Katzfuss, 2021), dynamic SPDE models (Cameletti et al., 2013; Krainski et al., 2019), and dynamic NNGPs (Datta et al., 2016b). Further gains in scalability were achieved through sparse Cholesky approximations (Jurek and Katzfuss, 2022) and variational inference (Hamelijnck et al., 2021). Simultaneously, a growing body of research has integrated neural networks with spatiotemporal models (Chen et al., 2024; Wikle and Zammit-Mangion, 2023). These developments have substantially broadened the range of spatiotemporal processes that can be modeled at scale.

However, substantial computational challenges remain in spatiotemporal settings because inference is still tied to covariance structures defined jointly over locations and time points. For example, to the best of our knowledge, the R package sdmTMB (Anderson et al., 2025) provides one of the fastest implementations of spatiotemporal models by exploiting the SPDE approach for spatial modeling and the template model builder (TMB)-based automatic differentiation for scalable inference (see Osgood-Zimmerman and Wakefield, 2023, for a review). However, as evaluated in Section 4.3, its spatiotemporal model still required 1416.3 s for panel data consisting of 800 spatial units observed over 200 time points. Given the recent increase in the spatial and temporal resolutions of spatiotemporal datasets, further improvements in scalability are required.

Coarse-to-fine spatial modeling (CF-SM) was recently introduced as an alternative framework for modeling spatial processes through the synthesis of local models rather than through explicit covariance modeling (Murakami et al., 2026a). The covariance-free framework achieves computational scalability while preserving predictive accuracy. Murakami et al. (2026b) extended this framework to spatial generalized linear mixed models. The present study builds directly on this formulation. However, both formulations were designed for spatial data and therefore do not account for temporal dynamics.

To address this limitation, we extended the CF-SM to spatiotemporal settings. The proposed method, termed coarse-to-fine spatiotemporal modeling (CF-STM), represents the underlying spatial process as a synthesis of multiscale local models, similar to CF-SM. At each spatial scale, temporal processes are modeled using state-space models. As will be shown later, CF-STM dramatically reduces the computation time relative to that of modern spatiotemporal statistical models while maintaining comparable predictive accuracy. For example, it took only 3.4 seconds for the aforementioned panel data (see Section 4.3).

The remainder of this paper is organized as follows: Section 2 describes CF-SM. Section 3 describes the development of CF-STM. Section 4 presents the Monte Carlo experiments. Section 5 reports an application to land price prediction, and Section 6 concludes the paper.

## 2. Coarse-to-fine spatial modeling (CF-SM)

Instead of modeling spatial dependence using a covariance function, as in Gaussian processes, CF-SM represents spatial processes by aggregating local models distributed across the study region. Let $z(s_i)$ denote a spatial process at location $s_i$ in a study region $\mathcal{D} \subset \mathbb{R}^2$. CF-SM represents $z(s_i)$ as the sum of $R$ scale-wise processes,

$$z(s_i) = \sum_{r=1}^{R} z_r(s_i), \tag{1}$$

where $z_r(s_i)$ denotes the single-scale process at scale *r*, with bandwidths recursively defined by $h_r = \alpha h_{r-1}$, where $\alpha = 0.9$ is assumed in this study. As the bandwidth decreases, the corresponding process captures increasingly finer spatial patterns. Thus, $z_r(s_i)$ represents the *r*-th largest-scale process, whereas $z_R(s_i)$ corresponds to the smallest-scale process. The number of scales, *R,* is considered unknown.

Each single-scale process, $z_r(s_i)$ is constructed by synthesizing local models distributed across the study region. Let $c_r \in \{1, \ldots, C_r\}$ index local centers at scale *r*. The local model associated with the center $c_r$ characterizes the spatial process in its neighborhood, and such local models are aggregated (see Section 3.3) to construct the single-scale process $z_r(s_i)$. By sequentially combining the resulting processes $z_1(s_i), \ldots, z_R(s_i)$ from coarse to fine scales, the CF-SM provides a flexible process $z(s_i)$ that captures spatial patterns across multiple scales.

This covariance-free formulation avoids heavy covariance matrix operations that often render conventional spatiotemporal models computationally prohibitive for large datasets. In addition, because CF-SM does not require an explicit likelihood evaluation, it can be extended to spatiotemporal modeling in a relatively straightforward and computationally efficient manner. For these reasons, this study extends the CF-SM to spatiotemporal settings.

## 3. Coarse-to-fine spatiotemporal modeling (CF-STM)

This section describes the development of CF-STM. For notational simplicity, we present the algorithm for regular panel data, where $y_t(s_i)$ is observed repeatedly at locations $s_i \in \{s_1, \ldots, s_{N_s}\}$ in a study region $\mathcal{D} \subset \mathbb{R}^2$ over time points $t \in \{1, \ldots, N_t\}$, yielding a sample size of $N = N_s N_t$. Nevertheless, CF-STM is readily applicable to irregularly observed data where observation sites may vary over time.

### 3.1. Data model

We consider the following generalized linear mixed model:

$$y_t(s_i) \sim F(\mu_t(s_i), \varphi), \qquad g\big(\mu_t(s_i)\big) = \mu_t^{lin}(s_i) = \mathbf{x}_t(s_i)'\boldsymbol{\beta} + \sum_{r=1}^{R} z_{t,r}(s_i) \tag{2}$$

where $F(\cdot, \varphi)$ denotes an exponential family distribution with dispersion parameter $\varphi$, and $g(\cdot)$ is a link function. For example, $F(\mu, \varphi) = N(\mu, \varphi)$ and $g(\mu) = \mu$ for Gaussian responses, whereas $F(\mu) = Poisson(\mu)$ and $g(\mu) = \log(\mu)$ for Poisson responses. $\mathbf{x}_t(s_i)$ denotes a vector of covariates, $\boldsymbol{\beta}$ is the corresponding coefficient vector, and $z_{t,r}(s_i)$ denotes the latent stochastic process defined at spatial scale *r* and time *t*.

Using a first-order Taylor expansion, Eq. (2) can be approximated using the following weighted linear model:

$$\eta_t(s_i) = \mathbf{x}_t(s_i)'\boldsymbol{\beta} + \sum_{r=1}^{R} z_{t,r}(s_i) + \varepsilon_t(s_i), \qquad E[\varepsilon_t(s_i)] = 0, \qquad V[\varepsilon_t(s_i)] = \frac{\sigma^2}{w_t(s_i)}, \tag{3}$$

which is based on working response $\eta_t(s_i) = \mu_t^{lin}(s_i) + \left(y_t(s_i) - \mu_t(s_i)\right)\frac{d\mu_t^{lin}(s_i)}{d\mu_t(s_i)}$ and weight $w_t(s_i) = \frac{1}{V(\mu_t(s_i))}\left(\frac{d\mu_t(s_i)}{d\mu_t^{lin}(s_i)}\right)^2$. In generalized linear models (GLMs), deviance loss is minimized by iteratively minimizing the weighted squared error. Similarly, we iteratively minimize the weighted squared error of Eq. (3) while sequentially increasing *R*. At each iteration, the local models for the working residuals are estimated and aggregated to construct the spatial process $z_{t,R}(s_i)$. This process is accepted only when it reduces deviance loss. Repeating this procedure sequentially learns the multiscale process $\sum_{r=1}^{R} z_{t,r}(s_i)$, which minimizes deviance loss.

Below, Section 3.2 describes the local model, Section 3.3 presents the aggregation of the local models, and Section 3.4 introduces the main algorithm that repeatedly minimizes the deviance loss and selects *R*. Subsequently, Section 3.5 quantifies uncertainty in the predictive values based on an empirically calibrated predictive variance. Finally, Section 3.6 discusses the spatiotemporal covariance structure implied by the proposed procedure.

### 3.2. Local model

Given the estimates $\hat{\boldsymbol{\beta}}$ and $\hat{z}_{t,1}(s_i), \dots, \hat{z}_{t,R-1}(s_i)$, the linearized model (Eq. 3) reduces to $\hat{\varepsilon}_{t,R-1}(s_i) \sim N\left(z_{t,R}(s_i), \frac{\sigma^2}{w_t(s_i)}\right)$ where $\hat{\varepsilon}_{t,R-1}(s_i) = \eta_t(s_i) - \mathbf{x}_t(s_i)'\hat{\boldsymbol{\beta}} - \sum_{r=1}^{R-1} \hat{z}_{t,r}(s_i)$ . To estimate the local mean $z_{t,c_R}$ and variance $\sigma^2_{t,c_R}$ at each local center $c_R \in \{1_R, \dots, C_R\}$, we geographically weight the reduced model as follows:

$$\hat{\varepsilon}_{t,R-1}(s_i) \mid c_R \sim N\left(z_{t,c_R}, \frac{\sigma^2_{t,c_R}}{k_{c_R}(s_i) w_t(s_i)}\right), \tag{4}$$

where $k_{c_R}(s_i)$ represents the kernel weight, which decreases with the Euclidean distance $d_{c_R}(s_i)$ between the *i*-th sample site and the $c_R$-th local center. Following CF-SM, we employ the exponential kernel $k_{c_R}(s_i) = \exp\left(-d_{c_R}(s_i)/h_r\right)$ where $h_r$ denotes the bandwidth associated with scale *r*. Estimating Eq. (4) using locally weighted least squares yields the local mean estimate $\hat{z}^{(s)}_{t,c_R} = E\left[z_{t,c_R}\right]$ together with its variance $\hat{\sigma}^{(s)2}_{t,c_R} = V\left[\hat{z}^{(s)}_{t,c_R}\right]$. The estimator can be represented as:

$$\hat{z}^{(s)}_{t,c_R} = z_{t,c_R} + e_{t,c_R},\ e_{t,c_R} \sim N\left(0, \hat{\sigma}^{(s)2}_{t,c_R}\right). \tag{5}$$

The local estimate $\hat{z}^{(s)}_{t,c_R}$ may be regarded as a noisy pseudo-observation of the latent process $z_{t,c_R}$, which is independently estimated at each time point using Eq. (4) and therefore ignores temporal dependence. With this in mind, we infer the temporal process underlying the sequence of pseudo-observations $\hat{z}^{(s)}_{1,c_R}, \dots, \hat{z}^{(s)}_{N_t,c_R}$ at each local center $c_R$, by fitting a state-space model that adopts Eq. (5) as the observation model, along with the following state-transition model:

$$z_{t,c_R} = f_{\boldsymbol{\theta}}\left(z_{1,c_R}, \dots, z_{t-1,c_R}\right) + u_{t,c_R}, \quad u_{t,c_R} \sim N(0, v^2), \tag{6}$$

where $v^2$ denotes a variance parameter. $f_{\boldsymbol{\theta}}(\cdot)$ denotes a linear state transition function, and $\boldsymbol{\theta}$ denotes its associated parameters. The function may be $f_\rho\left(z_{t-1,c_R}\right) = \rho z_{t-1,c_R}$ representing the first-order autoregressive model with temporal dependence parameter $\rho$, or other linear specifications accommodating higher-order dependence and seasonal dependence. A Kalman filter and smoother were used to obtain the filtered and smoothed distributions of $z_{t,c_R}$, respectively. In what follows, we use the latter, denoted by $z_{t,c_R} | \hat{z}^{(s)}_{1:N_t,c_R} \sim N\left(\hat{z}_{t,c_R}, \hat{\sigma}^2_{t,c_R}\right)$ where $\hat{z}_{t,c_R}$ and $\hat{\sigma}^2_{t,c_R}$ denote the smoothed mean and variance.

In summary, given $\widehat{\boldsymbol{\beta}}$ and $\hat{z}_{t,1}(s_i), \dots, \hat{z}_{t,R-1}(s_i)$, the local mean $\hat{z}^{(s)}_{t,c_R}$ and its variance $\hat{\sigma}^{(s)2}_{t,c_R}$ are estimated from Eq. (4) for each local center, $c_R$, and time step *t*. These quantities are then used to infer the temporally smoothed process $z_{t,c_R} | \hat{z}^{(s)}_{1:N_t,c_R} \sim N\left(\hat{z}_{t,c_R}, \hat{\sigma}^2_{t,c_R}\right)$ for each local center. Section 3.3 aggregates these temporal processes to obtain the single-scale process $z_{t,R}(\mathrm{s}_i)$.

### 3.3. Aggregation of the local models

The *R*-th single-scale process $z_{t,R}(s_i)$ is specified by spatially aggregating the temporally smoothed distributions $z_{t,c_R} | \hat{z}^{(s)}_{1:N_t,c_R} \sim N\left(\hat{z}_{t,c_R}, \hat{\sigma}^2_{t,c_R}\right)$ estimated at each local center $c_R$.

The aggregated model is constructed using the generalized product-of-experts method (Cao and Fleet, 2014) as follows:[1]

$$p\left(z_{t,R}(s_i)\right) = \prod_{c_R=1}^{C_R} p\left(z_{t,c_R} | \hat{z}_{1:N_t,c_R}^{(s)}\right)^{\tilde{k}_{c_R}(s_i)}. \tag{7}$$

Here, $\tilde{k}_{c_R}(s_i) = k_{c_R}(s_i) / \sum_{c_R} k_{c_R}(s_i)$ is the normalizing weight satisfying $\sum_{c_R} \tilde{k}_{c_R}(s_i) = 1$. Because $z_{t,c_R} | \hat{z}_{1:N_t,c_R}^{(s)}$ is Gaussian, Eq. (7) yields the following closed-form expression:

$$z_{t,R}(s_i) \sim N\left(\hat{z}_{t,R}(s_i), \hat{\sigma}_{t,R}^2(s_i)\right),$$

$$\hat{z}_{t,R}(s_i) = \hat{\sigma}_{t,R}^2(s_i) \sum_{c_R=1}^{C_R} \frac{\tilde{k}_{c_R}(s_i)}{\hat{\sigma}_{t,c_R}^2} \hat{z}_{t,c_R}, \qquad \hat{\sigma}_{t,R}^2(s_i) = \frac{1}{\sum_{c_R=1}^{C_R} \frac{\tilde{k}_{c_R}(s_i)}{\hat{\sigma}_{t,c_R}^2}}. \tag{8}$$

Repeating this reconstruction for all scales $r = 1, \ldots, R$ yields the multiscale spatiotemporal process:

$$\hat{z}_{t,1:R}(s_i) = \sum_{r=1}^{R} \hat{z}_{t,r}(s_i), \qquad \hat{\sigma}_{t,1:R}^2(s_i) = \sum_{r=1}^{R} \hat{\sigma}_{t,r}^2(s_i). \tag{9}$$

Although the predictive variance $\hat{\sigma}_{t,1:R}^2(s_i)$ is constructed assuming independence among scale-specific processes, this assumption may not hold exactly in practice. Therefore, the variance is calibrated as described in Section 3.5.

[1] This aggregation is optimal in the sense that the aggregated model minimizes the weighted average Kullback–Leibler divergence to each local model $z_{t,c_R} | \hat{z}_{1:N_t,c_R}^{(s)} \sim N\left(\hat{z}_{t,c_R}, \hat{\sigma}_{t,c_R}^2\right)$.

3.4. Learning algorithm

This section describes the sequential holdout validation (HV) algorithm used to select the number of scales $R$. During the HV, the $N_s$ sample sites are randomly divided into training sites and validation sites, with 75% used for training and 25% for validation. The set of validation sites is denoted as $V$. The algorithm selects $R$ by minimizing the validation deviance loss, defined as $Loss^{(Gau)} = \sum_{t=1}^{N_t} \sum_{i \in V} \left(y_t(s_i) - \hat{y}_t(s_i)\right)^2$ for Gaussian responses and $Loss^{(Poi)} = 2\sum_{t=1}^{N_t} \sum_{i \in V} \left( y_t(s_i) \log\left(\frac{y_t(s_i)}{\hat{y}_t(s_i)}\right) - \left(y_t(s_i) - \hat{y}_t(s_i)\right) \right)$ for Poisson responses.

The algorithm proceeds sequentially from coarse to fine scales. It starts with $R = 1$, where the predictive mean of the largest-scale process, $\hat{z}_1(s_i)$, is estimated using the training samples. This predictive mean is accepted if it reduces the validation loss; otherwise, it is set to zero. Next, $R = 2$ is considered and $\hat{z}_2(s_i)$ is estimated conditional on the previously accepted process, $\hat{z}_1(s_i)$. Again, $\hat{z}_2(s_i)$ is accepted only if it reduces validation loss. The multiscale process is then updated to $\hat{z}_{1:2}(s_i) = \hat{z}_1(s_i) + \hat{z}_2(s_i)$. This procedure is repeated by sequentially adding finer-scale components. At scale $R$, the candidate predictive mean $\hat{z}_R(s_i)$ is estimated conditional on the accumulated process, $\hat{z}_{1:R-1}(s_i) = \sum_{r=1}^{R-1} \hat{z}_r(s_i)$, and $\hat{z}_R(s_i)$ is retained only when the validation loss improves. In our implementation, the algorithm is terminated when the validation loss fails to improve for five consecutive values of $R$. The optimal number of scales is selected as the last value of $R$ that improved the validation loss.

More specifically, after initializing $R = 1$, setting $Loss_{R-1} = \infty$, $\hat{z}_{1:R-1}(s_i) = 0, Q = 0$, and obtaining the initial coefficient estimates $\widehat{\boldsymbol{\beta}}_R$ from the basic GLM, the algorithm proceeds as follows:

(i) Form $\hat{\mu}_t^{lin}(s_i) = \mathbf{x}_t(s_i)'\widehat{\boldsymbol{\beta}}_R$ , $\hat{\mu}_t(s_i) = g^{-1}\left(\hat{\mu}_t^{lin}(s_i)\right)$ , $\hat{\eta}_t(s_i) = \hat{\mu}_t^{lin}(s_i) + \left(y_t(s_i) - \hat{\mu}_t(s_i)\right)\frac{d\hat{\mu}_t^{lin}(s_i)}{d\hat{\mu}_t(s_i)}$, and $\widehat{w}_t(s_i) = \frac{1}{V(\hat{\mu}_t(s_i))}\left(\frac{d\hat{\mu}_t(s_i)}{d\hat{\mu}_t^{lin}(s_i)}\right)^2$.

(ii) Evaluate $\hat{\varepsilon}_{t,R-1}(s_i) = \hat{\eta}_t(s_i) - \mathbf{x}_t(s_i)'\widehat{\boldsymbol{\beta}}_R - \hat{z}_{1:R-1}(s_i)$.

(iii) Evaluate $\hat{z}_R(s_i)$ as follows:

(a) Distribute $C_R$ local centers throughout the study region. Following CF-SM, the centers were defined as *k*-means cluster centroids, where $C_R = \text{round}(1.5D^2/h_R^2)$. $D$ is the diagonal length of the bounding box enclosing all the sample sites.

(b) For each local center $c_R$ and time point *t*, fit the local model (Eq. 4) on $\hat{\varepsilon}_{t,R-1}(s_i)$ to estimate the local mean $\hat{z}_{t,c_R}^{(s)}$ and variance $\hat{\sigma}_{t,c_R}^{(s)2}$.

(c) For each center $c_R$, fit the state-space model (Eqs. 5–6) to the sequence $\hat{z}_{1,c_R}^{(s)}, \dots, \hat{z}_{N_t,c_R}^{(s)}$ and $\hat{\sigma}_{1,c_R}^{(s)2}, \dots, \hat{\sigma}_{N_t,c_R}^{(s)2}$ , and estimate the Kalman-smoothed distribution $z_{t,c_R}|\hat{z}_{1:N_t,c_R}^{(s)} \sim N\left(\hat{z}_{t,c_R}, \hat{\sigma}_{t,c_R}^2\right)$. If $R$ = 1, the parameters $\boldsymbol{\theta}$ and $v^2$ are estimated at this step by maximizing the joint likelihood across the $C_r$ local state space models. Estimating these parameters for a larger *R*, corresponding to finer minor components, resulted in unstable estimates and did not improve accuracy in our preliminary analysis.

Therefore, the parameters estimated at $R = 1$ are fixed and used across scales.

(d) For each time point $t$, the $C_R$ smoothed distributions are spatially aggregated using Eq. (8) to construct the single-scale process $z_{t,R}(s) \sim N\left(\hat{z}_{t,R}(s_i), \hat{\sigma}^2_{t,R}(s_i)\right)$.

(iv) Estimate $\widehat{\boldsymbol{\beta}}_R$ by minimizing the training deviance loss given $\hat{z}_{t,1:R-1}(s_i)$ and $\hat{z}_{t,R}(s)$. Iteratively reweighted least squares can be used, as in the basic GLM.

(v) Evaluate the validation loss $Loss_R$ for the model given the updated $\widehat{\boldsymbol{\beta}}_R$ and $\hat{z}_{t,R}(s_i)$:

(a) If $Loss_R < Loss_{R-1}$, the candidate scale is accepted. The parameters are updated as $\widehat{\boldsymbol{\beta}} = \widehat{\boldsymbol{\beta}}_R$ and $\hat{z}_{1:R}(s_i) = \hat{z}_{1:R-1}(s_i) + \hat{z}_R(s_i)$. Then, reset the counter $Q = 0$, and proceed to Step (vi).

(b) Otherwise, reject the candidate scale, update the counter as $Q \leftarrow Q + 1$, and retain the previous multiscale process: $\widehat{\boldsymbol{\beta}} = \widehat{\boldsymbol{\beta}}_{R-1}$ and $\hat{z}_{1:R}(s_i) = \hat{z}_{1:R-1}(s_i)$. If $Q$ is smaller than the threshold value, set to 5 in this study, proceed to Step (vi). Otherwise, the algorithm is terminated. The terminal resolution is $R$.

(vi) Update the scale as $R \leftarrow R + 1$, reduce the bandwidth according to $h_{R+1} = \alpha h_R$, where $\alpha = 0.9$ in this study, and return to Step (i).

In short, this algorithm sequentially estimates $\hat{z}_1(s_i), \ldots, \hat{z}_R(s_i)$ until the validation deviance loss ceases to improve. Because a candidate scale is accepted only when it reduces the validation loss, the algorithm does not increase the validation deviance loss over the iterations.

After the sequential HV, the same procedure without step (v) is repeated from $R = 1$ to the selected $R$ using the entire sample to obtain the final estimates $\hat{\boldsymbol{\beta}}$, $\hat{z}_{1:R}(s_i) = \sum_{r=1}^{R} \hat{z}_r(s_i)$, and $\hat{\sigma}_{t,1:R}^2(s_i) = \sum_{r=1}^{R} \hat{\sigma}_{t,R}^2(s_i)$ using Eq. (9).

3.5. Uncertainty quantification

3.5.1. Spatiotemporal process

As noted in Section 3.3, the predictive variance $\hat{\sigma}_{t,1:R}^2(s_i)$ may be biased because it is obtained under the assumption of independence across the scale-wise processes. Therefore, it is calibrated empirically using the variance estimated from the validation dataset. The empirical variance is given as $v_{raw}^{*2} = \max(v_{raw}^2, 0)$, where

$$v_{raw}^2 = \frac{\sum_t \sum_{i \in V} w_t(s_i) \left( \eta_t(s_i) - \mathbf{x}_t(s_i)' \hat{\boldsymbol{\beta}} - \hat{z}_{1:R}(s_i) \right)^2}{\sum_t \sum_{i \in V} w_t(s_i)} - \hat{\sigma}^2. \quad (10)$$

This subtraction follows from the decomposition of the residual variance, which is the first term on the right-hand side, into the latent process variance $v_{raw}^{*2}$ and the independent noise variance $\hat{\sigma}^2$. Conversely, the average of our predictive variance $\hat{\sigma}_{t,1:R}^2(s_i)$, which may be biased, is

$$v_{est}^2 = \frac{\sum_t \sum_{i \in V} w_t(s_i) \hat{\sigma}_{t,1:R}^2(s_i)}{\sum_t \sum_{i \in V} w_t(s_i)}. \quad (11)$$

We adjust the predictive variance $\hat{\sigma}_{t,1:R}^2(s_i)$ such that the average variance equals $v_{raw}^{*2}$ as follows:

$$\hat{\sigma}^{*2}_{t,1:R}(s_i) = \frac{v^{*2}_{raw}}{v^2_{est}} \hat{\sigma}^2_{t,1:R}(s_i). \tag{12}$$

The adjusted variance $\hat{\sigma}^{*2}_{t,1:R}(s_i)$ is used in the subsequent sections for uncertainty quantification. We emphasize that this adjustment is an empirical calibration procedure and a moment-matching step that aligns the average predictive variance with the validation-based estimate, rather than an exact statistical derivation. Because the calibration is global, it corrects the overall scale of the predictive variance but not its spatial or scale-wise allocation, as noted in Section 6.

### 3.5.2. Regression coefficients

The uncertainty in the regression coefficients was evaluated using a cluster-robust sandwich estimator (Bester et al., 2011), which accounts for residual dependence within spatial blocks and provides asymptotically valid standard errors. Let **X** denote the design matrix, **W** the diagonal matrix of the working weights, and let *g* index the spatial blocks. The covariance matrix of $\hat{\boldsymbol{\beta}}$ is estimated as

$$\hat{\boldsymbol{\Sigma}}_{\boldsymbol{\beta}} = (\mathbf{X}'\mathbf{W}\mathbf{X})^{-1}\left(\sum_g \mathbf{q}_g \mathbf{q}_g'\right)(\mathbf{X}'\mathbf{W}\mathbf{X})^{-1}, \tag{13}$$

where $\mathbf{q}_g = \sum_{(i,t)\in g} w_t(s_i)\big(\eta_t(s_i) - \mathbf{x}_t(s_i)'\hat{\boldsymbol{\beta}}\big)\mathbf{x}_t(s_i)$. For fast computation, Eq. (13) assumes dependence, characterized by $\mathbf{q}_g\mathbf{q}_g'$, only within each block.

The blocks were defined on a $G_x \times G_y$ spatial grid by cutting each coordinate at its empirical quantiles such that each block contained roughly the same number of sample sites. Based on a preliminary analysis, we set $G_x = \begin{cases} 2 & \text{if } G_x^0 < 2 \\ G_x^0 & \text{if } 2 \leq G_x^0 \leq 8 \\ 8 & \text{if } 8 < G_x^0 \end{cases}$ and defined $G_y$ analogously, where $G_x^0 = \text{round}(L_x / h^{med})$, $L_x$ and $L_y$ are the coordinate extents, and $h^{med} = \text{median}(h_1, \dots, h_R)$. This choice makes the average block side length approximately equal to $h^{med}$, such that that the blocks are reasonably large relative to the estimated dependence range. The lower and upper bounds (i.e., two and eight) prevent the number of blocks from becoming too small for stable estimation or too large, which would produce overly narrow blocks and leave residual dependence across neighboring blocks.

3.5.3. Response variables

For prediction at location $s_i$ and time *t*, the variance of the predictive mean $\hat{\mu}_t(s_i)$ is approximated using the delta method as

$$V[\hat{\mu}_t(s_i)] = \left(\frac{d\hat{\mu}_t(s_i)}{d\hat{\eta}_t(s_i)}\right)^2 V[\hat{\eta}_t(s_i)], \qquad V[\hat{\eta}_t(s_i)] = \mathbf{x}_t(s_i)' \hat{\boldsymbol{\Sigma}}_{\boldsymbol{\beta}} \mathbf{x}_t(s_i) + \hat{\sigma}_{t,1:R}^{*2}(s_i), \tag{14}$$

3.6. Property of the spatial process

CF-STM is best viewed as a prediction-oriented procedure that aggregates temporally smoothed local models rather than as a likelihood-based estimation of a single spatiotemporal model. Nevertheless, the resulting process implies a spatiotemporal covariance structure, which we characterize below.

Given the temporally smoothed local mean $\hat{z}_{t,c_R}$ and variance $\hat{\sigma}^2_{t,c_R}$, the predictive value at the $R$-th scale is expressed as $\hat{z}_{t,R}(s_i) = \sum_{c_R=1}^{C_R} w_{t,c_R}(s_i)\hat{z}_{t,c_R}$, where $w_{t,c_R}(s_i) = \frac{1}{Q_t(s_i)}\frac{k_{c_R}(s_i)}{\hat{\sigma}^2_{t,c_R}}$ is the $c_R$-th spatial kernel weighted by the temporal precision $1/\hat{\sigma}^2_{t,c_R}$, and $Q_t(s_i)$ ensures $\sum_{c_R=1}^{C_R} w_{t,c_R}(s_i) = 1$. This representation is analogous to process convolution and predictive process models (Higdon, 2002; Banerjee et al. 2008). The resulting covariance can be expressed as follows:

$$Cov\left[\hat{z}_{t,R}(s_i), \hat{z}_{t\prime,R}(s_i{}')\right] = \sum_{c_R=1}^{C_R} w_{t,c_R}(s_i) w_{t',c_R}(s_i{}') Cov\left[z_{t,c_R}, z_{t\prime,c_R}\right]. \qquad (15)$$

Because $w_{t,c_R}(s_i)$ varies over both space and time through $\hat{\sigma}^2_{t,c_R}$, the induced covariance structure is generally non-separable in space and time (Cressie and Huang, 1999; Gneiting, 2002).

Assuming independence across scales, the multiscale process $\hat{z}_{t,1:R}(s_i) = \sum_{r=1}^{R} \hat{z}_{t,r}(s_i)$ has covariance

$$Cov\left[\hat{z}_{t,1:R}(s_i), \hat{z}_{t\prime,1:R}(s_i{}')\right] = \sum_{r=1}^{R} Cov\left[\hat{z}_{t,r}(s_i), \hat{z}_{t\prime,r}(s_i{}')\right], \qquad (16)$$

which aggregates multiscale covariance functions, similar to fixed-rank kriging (Cressie and Johannesson, 2008) and multiresolution approximation (Katzfuss, 2017).

For example, suppose that the state transition model (Eq. 6) follows an autoregressive process, $z_{t,c_R} = \rho z_{t-1,c_R} + u_{t,c_R}, u_{t,c_R} \sim N(0, v^2)$ with $Cov[z_{t,c_R}, z_{t\prime,c_R}] = \frac{v^2}{1-\rho^2}\rho^{|t-t\prime|}$. Then, Eq. (15) reduces to $Cov[\hat{z}_{t,R}(s_i), \hat{z}_{t\prime,R}(s_i{}')] = \frac{v^2}{1-\rho^2}\rho^{|t-t'|}\kappa_{t,t',r}(s_i, s_i')$ where $\kappa_{t,t\prime,r}(s_i, s_i{}') = \sum_{c_R=1}^{C_R} w_{t,c_R}(s_i) w_{t',c_R}(s_i{}')$. Thus, the covariance is expressed as a temporally decaying function $\rho^{|t-t\prime|}$ multiplied by a time-dependent spatial kernel aggregation $\kappa_{t,t\prime,r}(s_i, s_i{}')$. It is separable only when $\kappa_{t,t\prime,r}(s_i, s_i{}')$ is constant over time. Appendix 3 further characterizes the covariance implied by CF-STM under an idealized process-convolution representation, where it becomes separable in space and time and, in a continuous-scale limit, reduces to a spatiotemporal Matérn covariance.

## 4. Monte Carlo experiments

### 4.1. Assumptions

This section compares the predictive accuracy and computation time of the proposed method for spatiotemporal data observed at locations $s_i$ randomly distributed within the region [0, 1] × [0, 1]. Here, we consider two data structures: regular panel data with $N_s = 400$ fixed sample sites and $N_t = 40$ time points, and irregularly observed data whose $400$ sample sites change randomly at every time point.

The synthetic data are generated using three response types: Gaussian, Poisson, and binomial:

$$\begin{aligned}
&\text{Gaussian: } y_t(\mathrm{s}_i) = \eta_t(s_i) + N(0, 0.5^2), \\
&\text{Poisson: } y_t(\mathrm{s}_i) \sim \mathrm{Pois}\big(e^{\eta_t(s_i)}\big), \qquad (17) \\
&\text{Binomial: } y_t(\mathrm{s}_i) \sim \mathrm{Bernoulli}\big(1/1 + e^{-\eta_t(s_i)}\big),
\end{aligned}$$

where $\eta_t(s_i) = \beta_0 + x_{1,t}(s_i)\beta_1 + x_{2,t}(s_i)\beta_2 + z_t(s_i)$. The spatiotemporal process $\mathbf{z}_t = \big[z_t(s_1), \dots, z_t(s_{N_s})\big]'$ is specified by the following temporal autoregressive process with spatially correlated innovations:

$$\mathbf{z}_t = \rho\, \mathbf{z}_{t-1} + \mathbf{W}\boldsymbol{\varepsilon}_t, \qquad \boldsymbol{\varepsilon}_t \sim N(\mathbf{0}, \mathbf{I}), \qquad (18)$$

initialized as $\mathbf{z}_1 = \frac{1}{\sqrt{1-\rho^2}} \mathbf{W}\boldsymbol{\varepsilon}_1$. The parameter $\rho$ determines the strength of the temporal dependence. **W** is the row-normalized spatial weight matrix whose ($i$, $j$)-th element equals $\tilde{k}(s_i, s_j) = \frac{1}{\sum_j \tilde{k}(s_i, s_j)} k(s_i, s_j)$ with $k(s_i, s_j) = \exp(-d(s_i, s_j)/0.1)$. The two covariates $x_{1,t}(s_i)$, $x_{2,t}(s_i)$ are generated in the same manner based on Eq. (18). Because each response distribution uses a different link function, the true coefficient values are specified separately for each distribution as follows:

$$\text{Gaussian: } (\beta_0, \beta_1, \beta_2) = (0.5,\ 2,\ -1.5),$$
$$\text{Poisson: } (\beta_0, \beta_1, \beta_2) = (0.5,\ 0.3,\ -0.2), \qquad (19)$$
$$\text{Binomial: } (\beta_0, \beta_1, \beta_2) = (0,\ 0.5,\ -0.4).$$

The coefficient magnitudes are set smaller for the Poisson and binomial specifications, which use log and logit links, respectively, such that the responses remain within an informative range.

We compare CF-STM with the following benchmark methods: a basic GLM; a GAM with a tensor-product smooth over space and time, built from a 40-basis spatial thin-plate spline and an 8-basis temporal spline (40 × 8 = 320 basis functions) (GAM); a generalized state space model with an AR(1) temporal process and a spatial process represented by a linear combination of 120 exponential kernels centered at k-means centroids with bandwidth $h = 0.1$ (GSSM); its multiscale extension using 40 kernels with $h = 0.2$, 80 kernels with $h = 0.1$, and 120 kernels with $h = 0.05$ (GSSM-MS); and a dynamic spatiotemporal model with an AR(1) temporal process and an SPDE-based spatial process (SPDE), whose spatial mesh is built with a cutoff (minimum node spacing) of 0.05; the sensitivity of the SPDE results to this choice is examined in Appendix 1. The R packages used to estimate the GAM, GSSM/GSSM-MS, and SPDE models are mgcv (Wood, 2017), KFAS (Helske, 2017), and sdmTMB (Anderson et al., 2025), respectively. Because the KFAS package does not accept regression coefficients fixed over time, the coefficients are estimated a priori by fitting the basic GLM, and the deviance residual is used to estimate the latent process.

Each model was trained using observations from a randomly selected 70% of the sample sites, and predictive accuracy was evaluated using observations from the remaining 30% of the sample sites. We used the root mean square error (RMSE), which quantifies the accuracy of the predictive mean, and the continuous ranked probability score (CRPS; Gneiting and Raftery, 2007), which quantifies the accuracy of the predictive distribution. For each combination of response type (Gaussian, Poisson, or binomial), strength of temporal dependence $\rho \in \{0.2, 0.7\}$, and data type (regular panel data or irregular data), RMSE and CRPS are evaluated over 200 repetitions, and their average values are reported.

All computations were performed using R 4.6.0 on a Mac Studio equipped with an Apple M3 Ultra chip and 512 GB of unified memory.

## 4.2. Results

Table 1 summarizes the predictive accuracies. The results for $\rho = 0.2$ and $\rho = 0.7$ are quite similar. CF-STM achieves a performance comparable to that of SPDE across all cases. Although the GSSM yields similarly small RMSE and CRPS values in the Gaussian case, CF-STM and SPDE produce substantially smaller RMSE and CRPS values in the Poisson and binomial cases.

These methods clearly outperform GLM, which ignores spatial and temporal dependence, and GAM.

These tendencies remained unchanged even for irregularly observed data. As summarized in Table 2, the differences in the RMSE between CF-STM and SPDE are consistently small, confirming the potential of CF-STM as an alternative to conventional covariance-based models, including SPDE.

Table 1: Predictive accuracy on the regular panel (bold: the first- and second-best methods).

| $\rho$ | Method | Gaussian | | Poisson | | Binomial | |
|---|---|---|---|---|---|---|---|
| | | RMSE | CRPS | RMSE | CRPS | RMSE | CRPS |
| 0.2 | GLM | 1.132 | 0.638 | 2.417 | 1.189 | 0.486 | 0.236 |
| | GAM | 1.012 | 0.571 | 2.194 | 1.089 | 0.475 | 0.226 |
| | GSSM | 0.588 | **0.335** | 1.753 | 0.858 | 0.461 | 0.212 |
| | GSSM-MS | 0.593 | 0.340 | 1.762 | 0.861 | 0.461 | 0.213 |
| | SPDE | **0.585** | **0.335** | **1.663** | **0.826** | **0.447** | **0.200** |
| | CF-STM | **0.584** | 0.340 | **1.667** | **0.829** | **0.448** | **0.201** |
| 0.7 | GLM | 1.472 | 0.831 | 2.413 | 1.194 | 0.485 | 0.235 |
| | GAM | 1.004 | 0.566 | 1.928 | 0.950 | 0.456 | 0.208 |
| | GSSM | 0.618 | **0.353** | 1.731 | 0.849 | 0.459 | 0.210 |
| | GSSM-MS | 0.627 | 0.359 | 1.740 | 0.852 | 0.459 | 0.211 |
| | SPDE | **0.613** | **0.352** | **1.651** | **0.819** | **0.444** | **0.197** |
| | CF-STM | **0.609** | 0.359 | **1.661** | **0.826** | **0.445** | **0.198** |

Table 2: RMSE values of SPDE and CF-STM across cases.

| Distribution | Method | Regular panel | | Irregular | |
|---|---|---|---|---|---|
| | | $\rho = 0.7$ | $\rho = 0.2$ | $\rho = 0.7$ | $\rho = 0.2$ |
| Gaussian | SPDE | 0.613 | 0.585 | 0.526 | 0.525 |
| | CF-STM | 0.609 | 0.584 | 0.528 | 0.524 |
| Poisson | SPDE | 1.651 | 1.663 | 1.564 | 1.562 |
| | CF-STM | 1.661 | 1.667 | 1.571 | 1.565 |
| Binomial | SPDE | 0.444 | 0.447 | 0.436 | 0.438 |
| | CF-STM | 0.445 | 0.448 | 0.437 | 0.438 |

Table 3 compares the bias and standard deviation of the estimated $\hat{\beta}_1$ values. The GSSM and GSSM-MS were excluded because the coefficients were estimated a priori using the standard GLM (see Section 4.1). In the Gaussian and Poisson cases, all spatiotemporal models successfully reduced both the bias and SD relative to the GLM. In particular, SPDE produced the smallest values in most cases. CF-STM also yielded a smaller bias and SD than GLM and GAM, with performance approaching that of SPDE. These results indicate that CF-STM clearly improves the coefficient estimation accuracy relative to GLM. In contrast, the binomial case suggests that there is room for further improvement, as the estimation bias of CF-STM is larger than that of SPDE and is comparable to that of GAM.

Table 3: Estimation accuracy of $\hat{\beta}_1$ on the regular panel ($\rho = 0.7$).

| Method | Gaussian | | Poisson | | Binomial | |
|---|---|---|---|---|---|---|
| | Bias | Std.dev. | Bias | Std.dev. | Bias | Std.dev. |
| GLM | $-0.019$ | $0.064$ | $-0.007$ | $0.036$ | $-0.120$ | $0.049$ |
| GAM | $0.009$ | $0.044$ | $0.003$ | $0.025$ | $-0.052$ | $0.042$ |
| SPDE | $0.001$ | $0.018$ | $-0.002$ | $0.015$ | $-0.003$ | $0.034$ |
| CF-STM | $-0.006$ | $0.029$ | $-0.001$ | $0.022$ | $-0.052$ | $0.036$ |

Table 4 reports the mean standard error (SE) of $\hat{\beta}_1$ and its empirical 95% coverage. While the SEs of GLM and GAM are far too small to account for spatiotemporal dependence (coverage 0.045 – 0.335), the SE of CF-STM achieves near-nominal coverage (0.875–0.970), on par with SPDE, confirming the reliability of its coefficient confidence intervals.

Table 4: Accuracy of the SEs of $\hat{\beta}_1$ on the regular panel ($\rho = 0.7$).

| Method | Gaussian | | Poisson | | Binomial | |
|---|---|---|---|---|---|---|
| | Mean SE | 95% coverage | Mean SE | 95% coverage | Mean SE | 95% coverage |
| GLM | 0.012 | 0.270 | 0.005 | 0.185 | 0.017 | 0.045 |
| GAM | 0.009 | 0.275 | 0.006 | 0.265 | 0.020 | 0.335 |
| SPDE | 0.016 | 0.935 | 0.013 | 0.890 | 0.032 | 0.930 |
| CF-STM | 0.027 | 0.925 | 0.031 | 0.970 | 0.037 | 0.875 |

## 4.3.Computation time

Table 5 presents a comparison of the computation times for the regular panel data. CF-STM was one to three orders of magnitude faster than GAM, GSSM, and SPDE across all sample sizes. For example, with $N_t = 40$ and $N_s = 400$, CF-STM took only 0.5 seconds, compared with 127.6 seconds for SPDE, 52.8 seconds for GAM, and 272.8 seconds for the three-scale GSSM-MS. With a larger sample of $N_t = 200$ and $N_s = 800$, the gap widens further, with CF-STM requiring only 3.4 seconds compared with 1,416.3 seconds for SPDE. The computational cost of SPDE depends on the mesh resolution, but refining the mesh does not close the accuracy gap in favor of CF-STM: as shown in Appendix 1, CF-STM matches the predictive accuracy of the finest SPDE meshes at a small fraction of their computation time.

Table 6 evaluates the computation time of CF-STM alone for larger samples, where other spatiotemporal models are difficult to apply. The computational cost increases linearly with both $N_s$ and $N_t$. Even in the case of $N = 2{,}500{,}000\ (= 500 \times 5000)$ samples, it took only 59.1 seconds for parameter estimation and 12.1 seconds for prediction at an equal number of locations. These results clearly indicate the scalability of CF-STM.

Table 5: Comparison of computation time for regular panel data (seconds).

| $N$ | $N_s$ | $N_t$ | GLM | GAM | GSSM | GSSM-MS | SPDE | CF-STM |
|---|---|---|---|---|---|---|---|---|
| 16,000 | | 40 | 0.00 | 52.8 | 34.5 | 272.8 | 127.6 | 0.5 |
| 36,000 | 400 | 80 | 0.01 | 105.6 | 65.9 | 545.4 | 300.1 | 0.6 |
| 80,000 | | 200 | 0.01 | 259.6 | 171.0 | 1363.2 | 904.1 | 1.4 |
| 32,000 | | 40 | 0.01 | 105.8 | 66.9 | 522.3 | 197.0 | 0.9 |
| 64,000 | 800 | 80 | 0.01 | 230.4 | 131.9 | 1089.1 | 470.2 | 1.5 |
| 160,000 | | 200 | 0.04 | 575.0 | 331.4 | 2618.0 | 1416.3 | 3.4 |

Table 6: Computation time of CF-STM for regular panel data (seconds).

| $N_t$ | $N_s$ | | | | | | | |
|---|---|---|---|---|---|---|---|---|
| | Model estimation | | | | Prediction | | | |
| | 400 | 800 | 2000 | 5000 | 400 | 800 | 2000 | 5000 |
| 40 | 0.4 | 0.7 | 2.2 | 6.6 | 0.1 | 0.2 | 0.6 | 1.4 |
| 80 | 0.5 | 1.2 | 3.6 | 11.0 | 0.1 | 0.3 | 0.8 | 2.3 |
| 200 | 1.1 | 2.5 | 7.8 | 23.9 | 0.3 | 0.6 | 1.6 | 4.8 |
| 500 | 2.8 | 6.5 | 19.5 | 59.1 | 0.6 | 1.3 | 3.7 | 12.1 |

In an additional experiment, we evaluated the computation time of CF-STM for irregularly observed data. We considered three cases with $N_t = 200$ and $N_s \in \{100, 400, 1000\}$, where the sample locations were regenerated at every time point. The sample sizes, which correspond to the number of unique sample locations, were 20,000, 80,000, and 200,000. The average computation time (estimation + prediction) over the five trials increased approximately linearly, taking 35.9, 151.2, and 405.5 seconds, respectively. Thus, CF-STM is computationally efficient not only for regular panel data but also for irregular data. Together with the results in Table 6, this finding suggests that the computation time of CF-STM depends strongly on the number of unique sample locations in the dataset, rather than *N*.

4.4.Robustness to a Gaussian-process-based data-generating process

To assess whether the preceding findings depend on the choice of data-generating process (DGP), we repeated the predictive accuracy and coefficient estimation experiments using a Gaussian-process-based DGP consistent with SPDE. Specifically, the spatial moving-average process $\mathbf{W}\boldsymbol{\varepsilon}_t$ in Eq. (18) was replaced with a zero-mean Gaussian process with a Matérn covariance (smoothness 1 and marginal variance 1), and the resulting field evolved according to the same temporal AR(1) process. The regression coefficients, training-test split, and evaluation metrics were unchanged. Full details and results are reported in Appendix 2.

CF-STM and SPDE remained essentially tied in predictive accuracy even under this SPDE-compatible DGP. For example, when temporal dependence was strong ($\rho$ = 0.7), their Gaussian RMSE values were 0.735 for CF-STM and 0.755 for SPDE, respectively. SPDE achieved a slightly smaller Gaussian CRPS; 0.451 for CF-STM and 0.435 for SPDE. Differences were also small for the Poisson and binomial responses. For coefficient estimation, SPDE attained the smallest bias and standard deviation overall, while CF-STM remained close behind and clearly outperformed GLM and GAM. The complete results in Tables A2 and A3 confirm that the conclusions of the main experiments are robust to the choice of DGP.

## 5. Application to land price analysis

### 5.1. Assumptions

This section applies CF-STM to an analysis of the officially assessed residential land prices in the Tokyo metropolitan area between 1984 and 2016 (JPY/m$^2$; National Land Numerical Information download site: https://nlftp.mlit.go.jp/ksj/). The irregularly observed dataset comprises 27,499 unique sites over 33 years ($N$ = 267,907). During this period, the Tokyo metropolitan area experienced suburban expansion in the 1980s and the early 1990s, a prolonged decline in land prices after the collapse of the asset price bubble, and a renewed concentration of land price growth in central Tokyo and selected suburban centers after the 2000s (see Figures 1 and 2). Against this background, this analysis aims to model the spatiotemporal evolution of residential land prices and identify the factors underlying these changes.

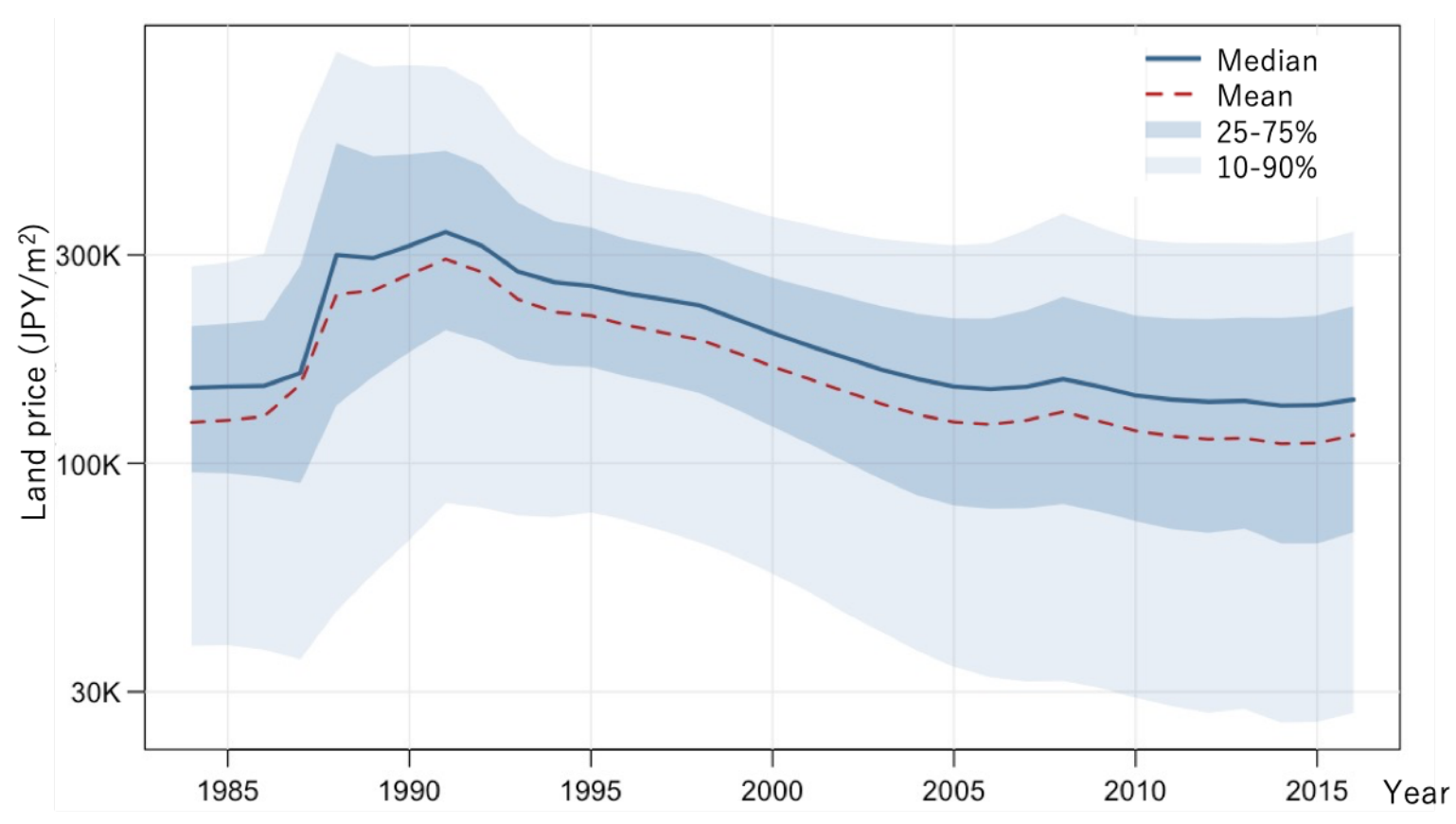

Figure 1: Temporal pattern of the land prices.

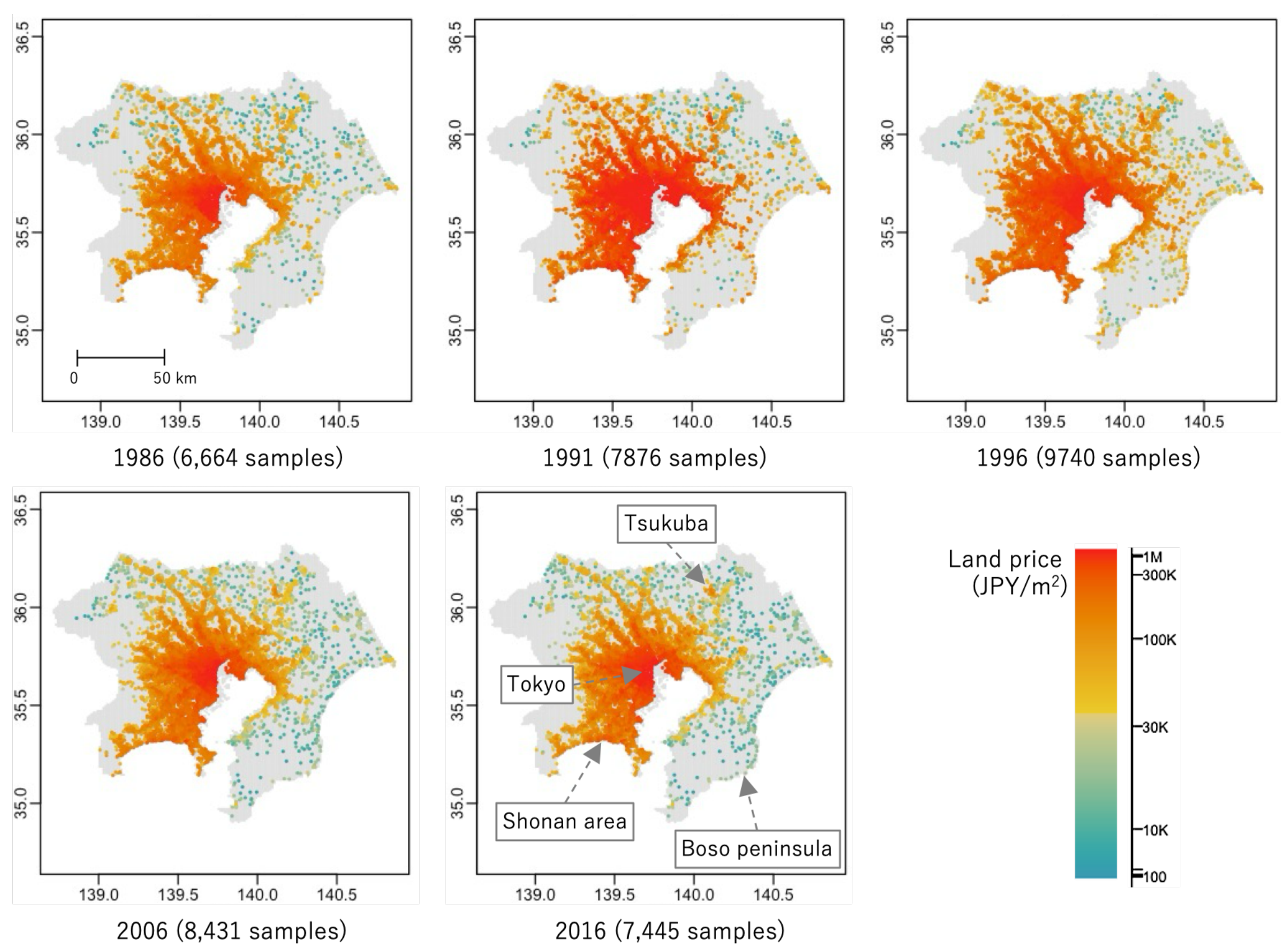

Figure 2: Residential land price data.

A Gaussian CF-STM was used as the main model for the analysis. In addition, a CF-STM with a Gamma response distribution (Gamma CF-STM) was applied to examine the practical usefulness of the non-Gaussian extension. The Gamma response distribution is suitable for positive continuous variables, such as land prices.

The response variables were the logged residential land price for the Gaussian CF-STM and the raw land price for the Gamma CF-STM. The covariates were the Euclidean distance from the nearest railway station (Station_dist; km), railway network distance from the nearest station to Tokyo station (Tokyo_dist; km), and the proportions of agricultural land, forest, wasteland, and water bodies (river, lake, and ocean) in 500-m grids, including the sample site (Agri, Forest, Waste, and Water).

The Gaussian and Gamma CF-STMs were used to predict annual land prices for each grid cell. The study area consisted of 14,742 grid cells with a spatial resolution of 1 km, and the spatial coordinates of each cell were defined by its geometric center.

### 5.2. Parameter estimation results

Table 7 reports the estimated parameters. For comparison, the basic linear regression model (LM) for the logged land prices was also estimated. The Gaussian CF-STM took 110.6 seconds for the estimation and prediction, while the Gamma CF-STM took 181.6 seconds. Both models indicated strong temporal dependence ($\rho$ = 0.951 and 0.922) and spatial dependence whose bandwidth ranges from 0.3 km to 80 km, capturing both large- and small-scale patterns. The validation RMSE, R2, and CRPS of the Gaussian and Gamma CF-STMs are substantially better than those of LM, confirming the flexibility of CF-STM in capturing latent spatiotemporal processes. The accuracies of the Gaussian and Gamma specifications are quite close, although the former is slightly better.

Table 7: Parameter estimation result (CI: Confidence Interval). The process noise SD, $v$, represents the AR(1) state-innovation SD in the state space model, and the data noise SD, $\sigma$, represents the SD of the residuals.

| | LM | | Gaussian CF-STM | | Gamma CF-STM | |
|---|---|---|---|---|---|---|
| | Est. | 95% CI | Est. | 95% CI | Est. | 95% CI |
| Intercept | 13.268 | [13.264, 13.272] | 13.253 | [13.079, 13.427] | 13.206 | [13.064, 13.347] |
| Tokyo_dist | −0.030 | [−0.030, −0.030] | −0.032 | [−0.039, −0.024] | −0.030 | [−0.036, −0.024] |
| Station_dist | −0.159 | [−0.160, −0.157] | −0.189 | [−0.242, −0.135] | −0.184 | [−0.224, −0.144] |
| Agri | −0.904 | [−0.916, −0.891] | −0.425 | [−0.779, −0.071] | −0.420 | [−0.691, −0.150] |
| Forest | −0.356 | [−0.373, −0.340] | −0.249 | [−0.690, 0.192] | −0.248 | [−0.601, 0.105] |
| Wasteland | −0.265 | [−0.340, −0.190] | −0.108 | [−0.998, 0.783] | −0.100 | [−0.906, 0.705] |
| Water | −0.565 | [−0.591, −0.538] | −0.232 | [−0.617, 0.153] | −0.231 | [−0.561, 0.100] |
| Num. of scales $R$ | | | 55 ($b_h$:80.2 km–0.33 km) | | 53 ($b_h$:80.2 km–0.27 km) | |
| Temporal corr. $\rho$ | | | 0.951 | | 0.922 | |
| Proc. noise SD $v$ | | | 0.321 | | 0.180 | |
| Data noise SD $\sigma$ | | | 0.063 | | 0.093 | |
| Valid. RMSE | 0.552 | | 0.181 | | 0.185 | |
| Valid. $R^2$ | 0.658 | | 0.962 | | 0.951 | |

| Valid. CRPS | 0.313 | 0.096 | 0.102 |
|---|---|---|---|

The LM and CF-STMs agreed on the sign of each coefficient. All the models indicate that land prices are higher in central Tokyo and near railway stations, confirming the importance of railway accessibility in the Tokyo metropolitan area. The LM suggests statistically significant negative effects for all natural land-use variables. In contrast, both the Gaussian and Gamma CF-STMs suggest weaker effects. Although Agri, Forest, Wasteland, and Water all had negative point estimates, their 95% confidence intervals included zero, except for Agri, whose interval only marginally excluded zero.

The magnitudes of the estimated coefficients differed between the LM and the CF-STMs. Compared with the LM, the CF-STMs shrank the land-use-related coefficients toward zero, whereas the absolute magnitudes of Tokyo_dist and Station_dist increased. Given the simulation results showing that CF-STM improves the coefficient estimation accuracy relative to GLM (Tables 3 and 4), these differences in the point and interval estimates may reflect omitted variable bias in the LM coefficients caused by the failure to account for latent spatiotemporal processes. These findings suggest that CF-STM is useful not only for prediction but also for regression analysis.

Figure 3 plots the predictive means of the scale-wise processes in 2001 extracted from the Gaussian CF-STM. The results suggest that land prices exhibit multiscale spatial patterns, ranging from a broad west-high/east-low pattern at a scale of several tens of kilometers to local patterns at scales of several kilometers.

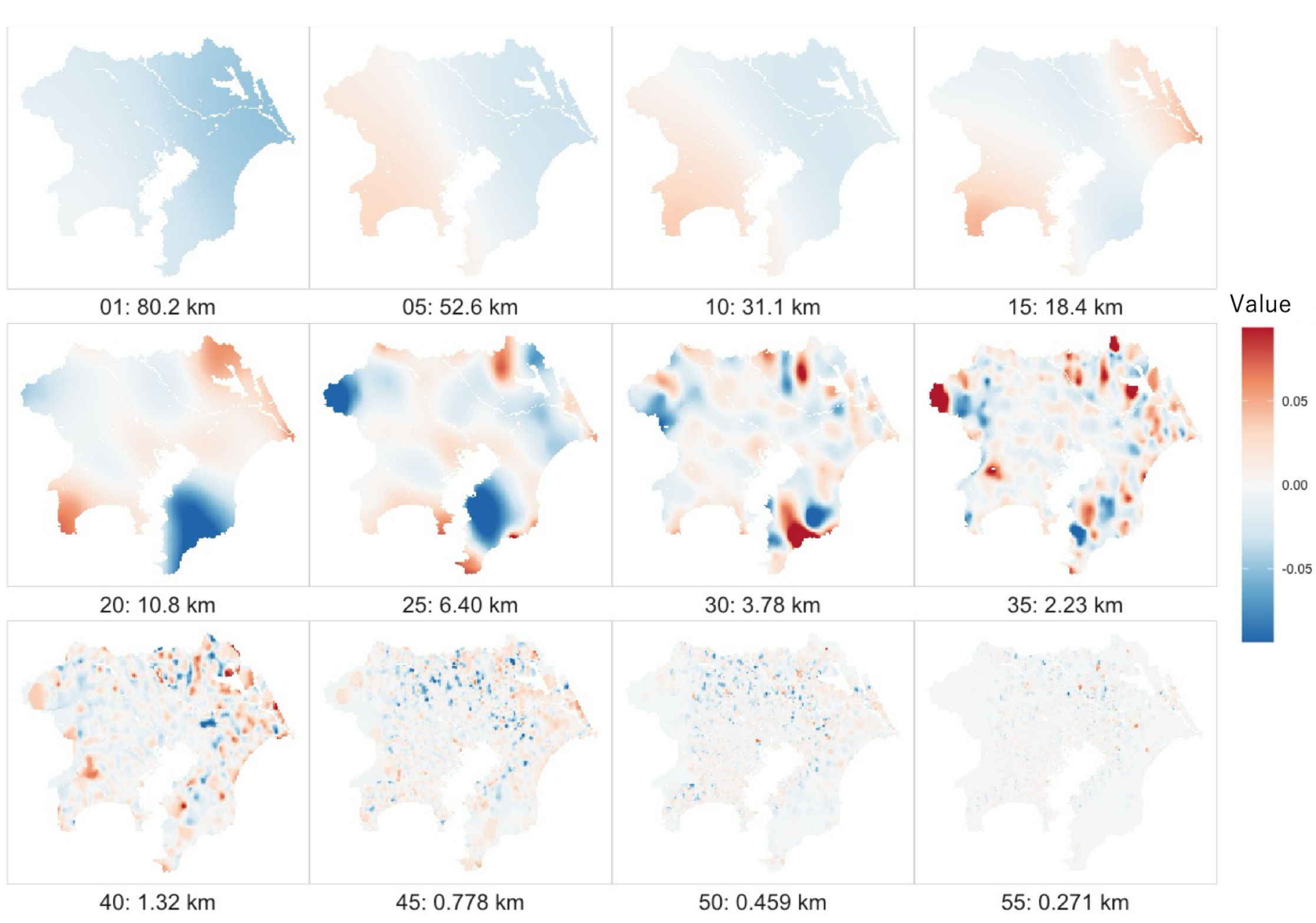

Figure 3: Predictive means of the scale-wise processes in 2001 and their bandwidth values. Due to space limitations, only the first and every fifth process are shown.

### 5.3. Prediction results

Figure 4 maps the land prices predicted by the Gaussian CF-STM. From 1986 to the peak of the bubble era in 1991, high-price areas expanded across the study region. Thereafter, suburban residential land prices declined until approximately 2006, strengthening the concentration of high prices in central Tokyo. By 2016, the high-price areas largely coincided with the railway commuting zone to central Tokyo. High land prices remained on the western side of central Tokyo, where railway services have been improved. In contrast, on the Boso Peninsula (Figure 2), where high-price areas were continuously distributed along the coast in 1991, most areas shifted to low-price zones after 2006, except for the northwestern region, which is relatively close to central Tokyo. Given that railway commuting from the southern Boso Peninsula to central Tokyo is impractical because of the long travel time, the decline in land prices in this area is likely related to poor accessibility to central Tokyo.

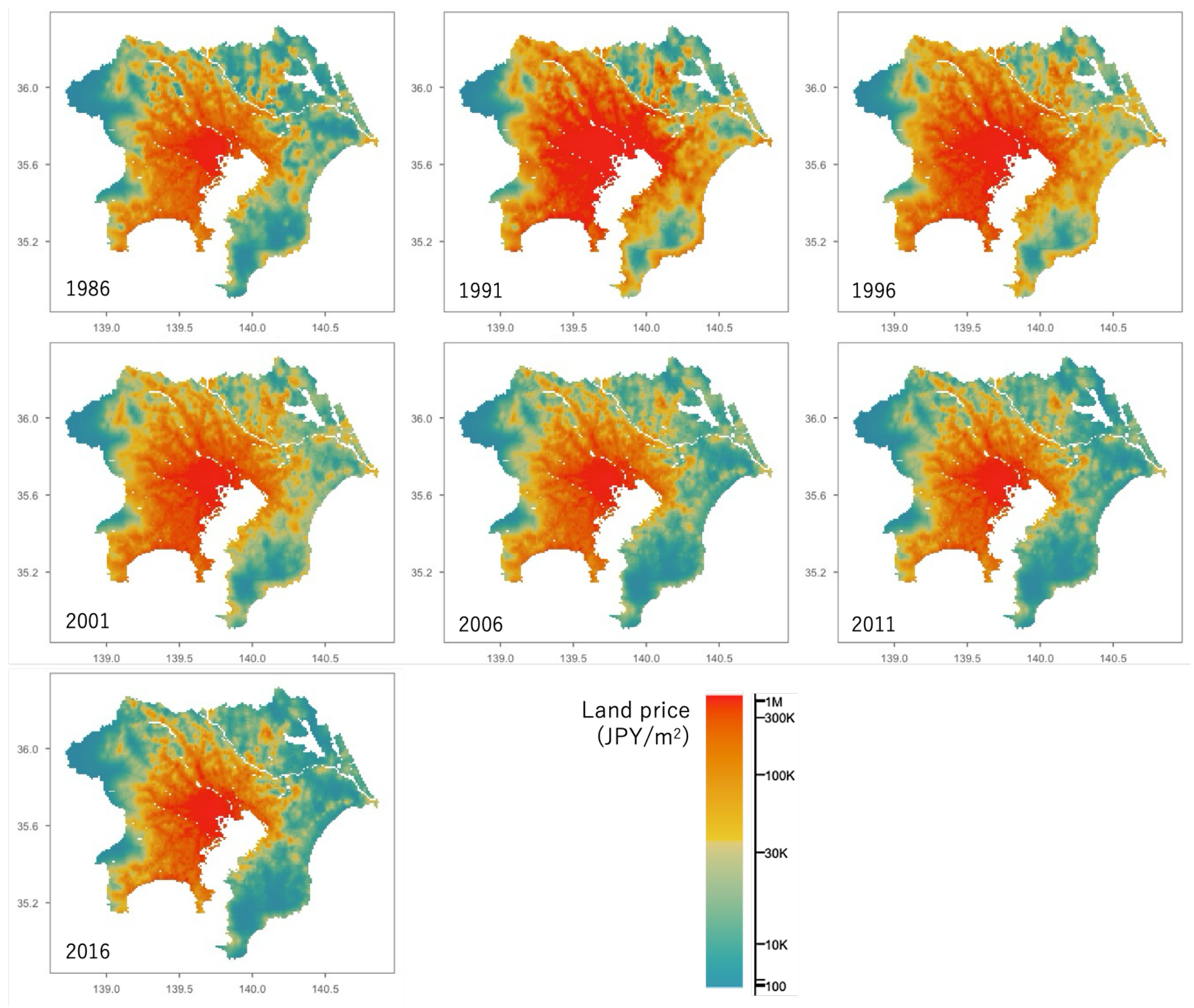

Figure 4: Predicted land prices (Gaussian CF-STM; 1986–2016).

Figure 5 maps the prediction results of Gamma CF-STM for 1986, 1991, and 1996. The predicted values were visually indistinguishable from those of the Gaussian CF-STM, suggesting the robustness of the interpolation results.

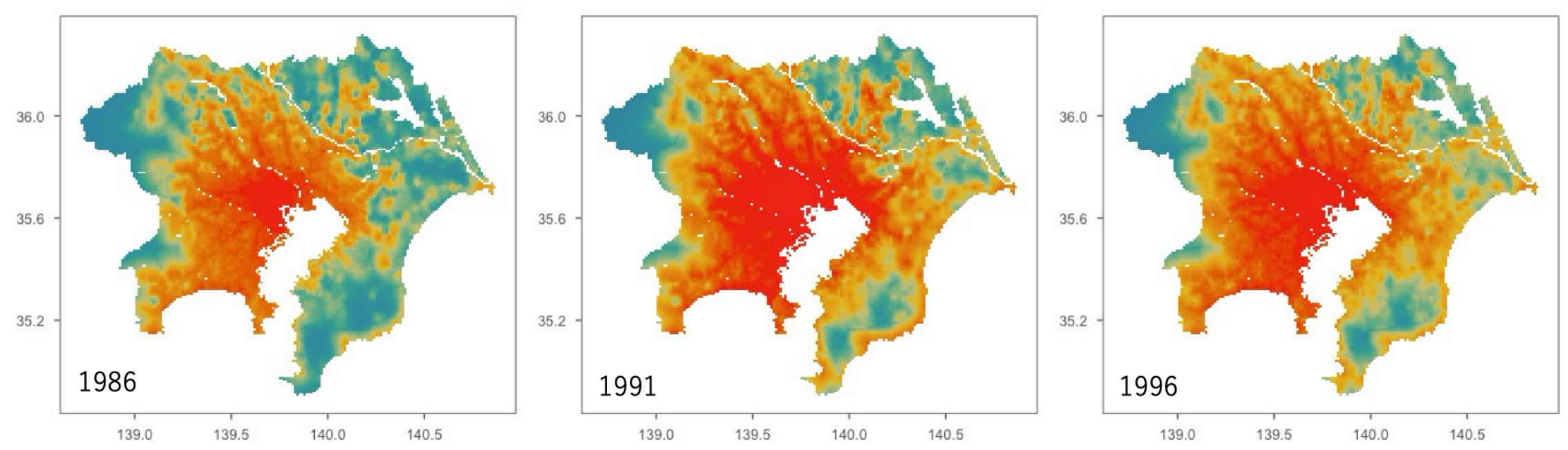

Figure 5: Predicted land prices (Gamma CF-STM; 1986–1996). See Figure 4 for the color legend.

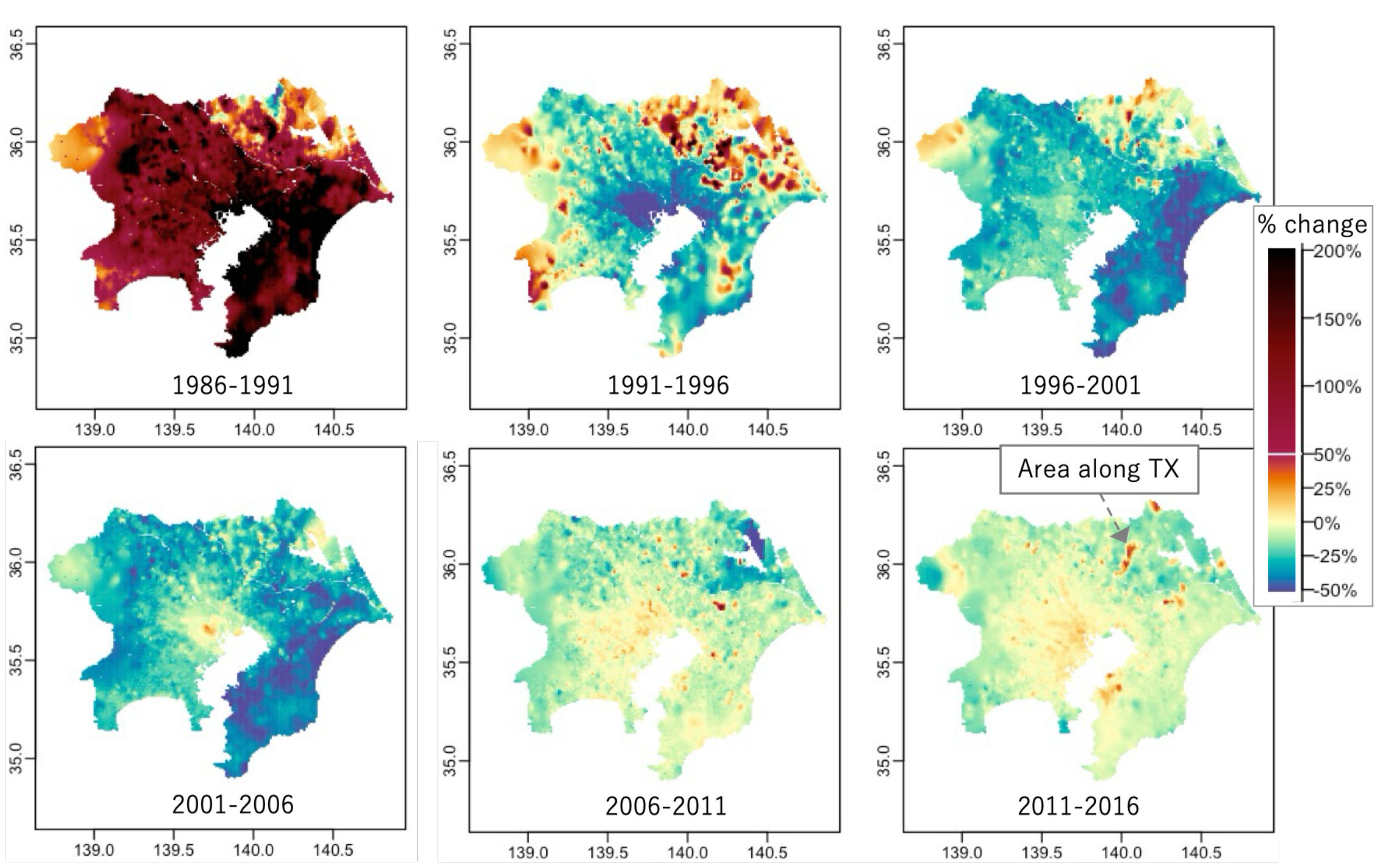

Figure 6: Predicted change rate of the land prices (CF-STM).

Figure 6 maps the five-year percentage changes in land prices predicted by the Gaussian CF-STM. The model successfully recovered the trajectory of the asset-price bubble, with rapid and widespread price increases during 1986–1991, followed by a persistent decline during 1991–2006 and stabilization after 2006. After 1991, land prices declined sharply in central Tokyo, whereas several peripheral areas, particularly in the northeastern suburbs, continued to experience price appreciation, suggesting a delayed collapse of the bubble. During 1996–2006, the land price decline became particularly pronounced in the Boso Peninsula, whereas renewed appreciation emerged in central Tokyo after the early 2000s. The 2011–2016 period further revealed substantial price increases along the Tsukuba Express (TX) line (Figure 6), a train line that opened in 2005, suggesting a strong influence of transport infrastructure improvements on local land markets. These results demonstrate that CF-STM successfully captured fine-scale spatial heterogeneity and its temporal transitions.

Uncertainty in the predictive values was quantified using the coefficient of variation, which was calculated as (predictive SD)/(predictive mean). Figures 7 and 8 show the predictive coefficients of variation obtained from the Gaussian and Gamma CF-STMs, respectively, for 1991 and 2016. Because the Gaussian CF-STM returns SDs on a log scale, while the Gamma CF-STM returns SDs on a real scale, these values are not directly comparable. Nevertheless, both indicated consistently lower predictive uncertainty in urban areas with dense observations and higher uncertainty in suburban areas where data coverage is limited. Although the uncertainty became slightly larger in 2016, when there were fewer observations (Figure 2), the overall spatial pattern remained largely unchanged over time.

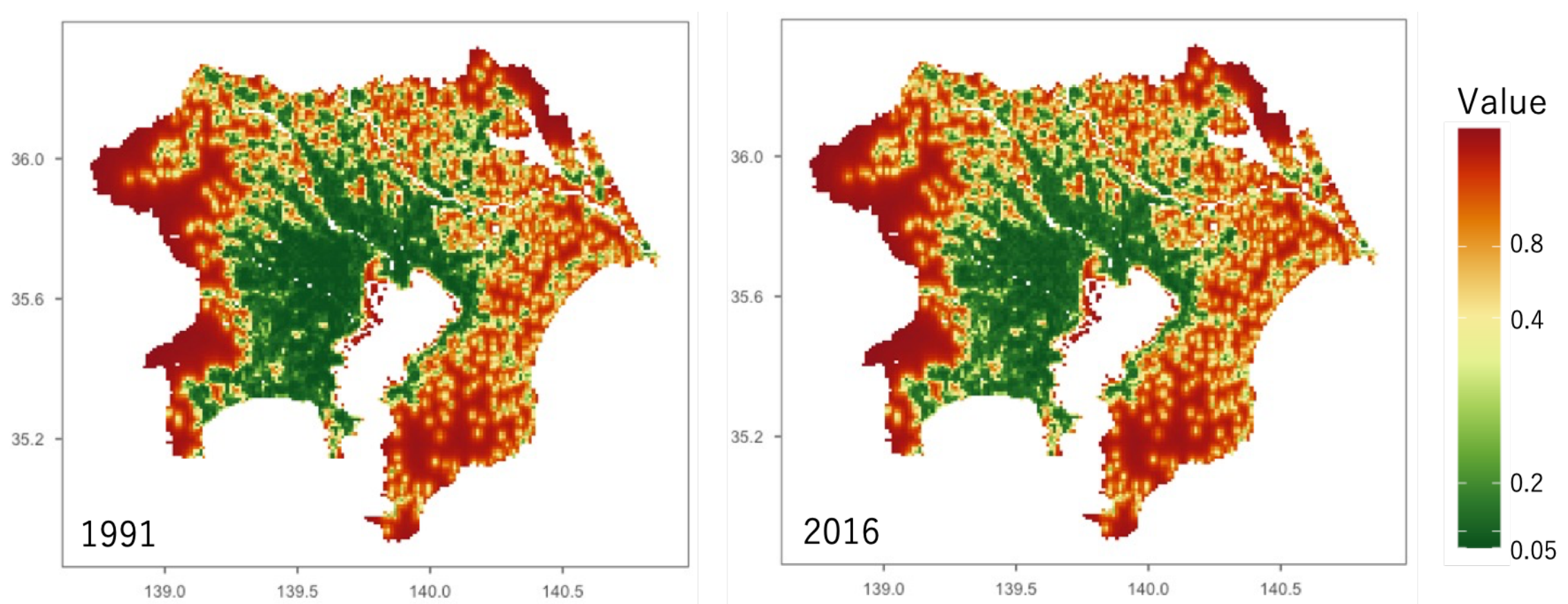

Figure 7: Predictive coefficient of variation (Gaussian CF-STM; log-scale).

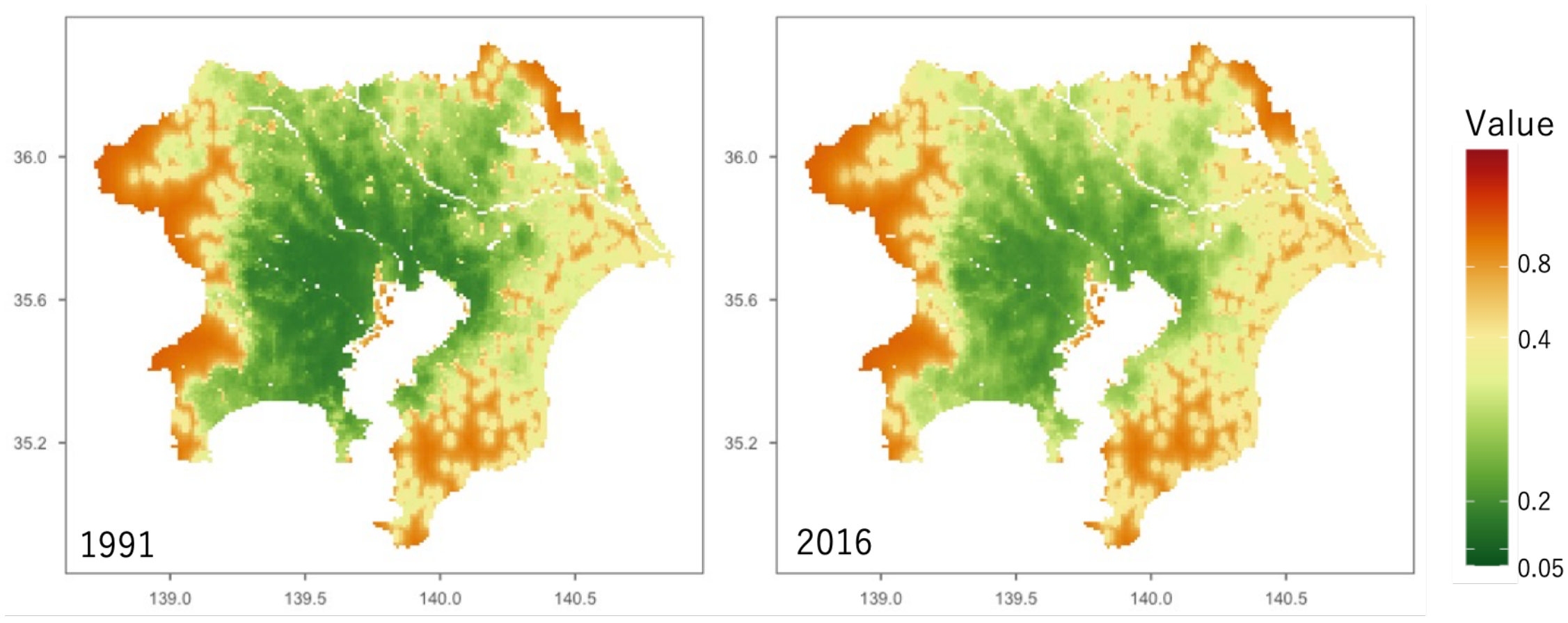

Figure 8: Predictive coefficient of variation (Gamma CF-STM; real-scale).

### 5.4. Predictive accuracy

This section compares the out-of-sample predictive accuracies of the LM, Gaussian CF-STM, Gamma CF-STM, and annual Gaussian CF-SM, which fit CF-SM independently for each year and therefore ignore temporal dependence. Prediction accuracy was evaluated using five-fold cross-validation, where the sample sites were randomly partitioned into five groups. In each fold, one group was held out, and the prediction accuracy was evaluated using this holdout group, which was excluded from the model estimation. The RMSE of LM was 0.559. The annual CF-SM substantially reduced this value to 0.208 by incorporating latent spatial processes, and the Gaussian CF-STM further reduced the RMSE to 0.189 by additionally modeling temporal dependence, whereas the RMSE of Gamma CF-STM was 0.194.

Figure 9 presents a comparison of the annual RMSE values. The CF-STMs tended to achieve lower RMSE values than the annual CF-SM during the bubble period (1987–1992), demonstrating the importance of accounting for temporal dependence when capturing the rapid fluctuations in land prices during this period. Moreover, CF-STM remained consistently more accurate after the early 2000s, when temporal changes became relatively gradual. Notably, the annual Gaussian CF-SM required 1,485 s to estimate separate spatial models for each year, whereas the Gaussian CF-STM took only 110.6 s. These results indicate that the CF-STMs outperform year-by-year spatial modeling in terms of both predictive accuracy and computational efficiency.

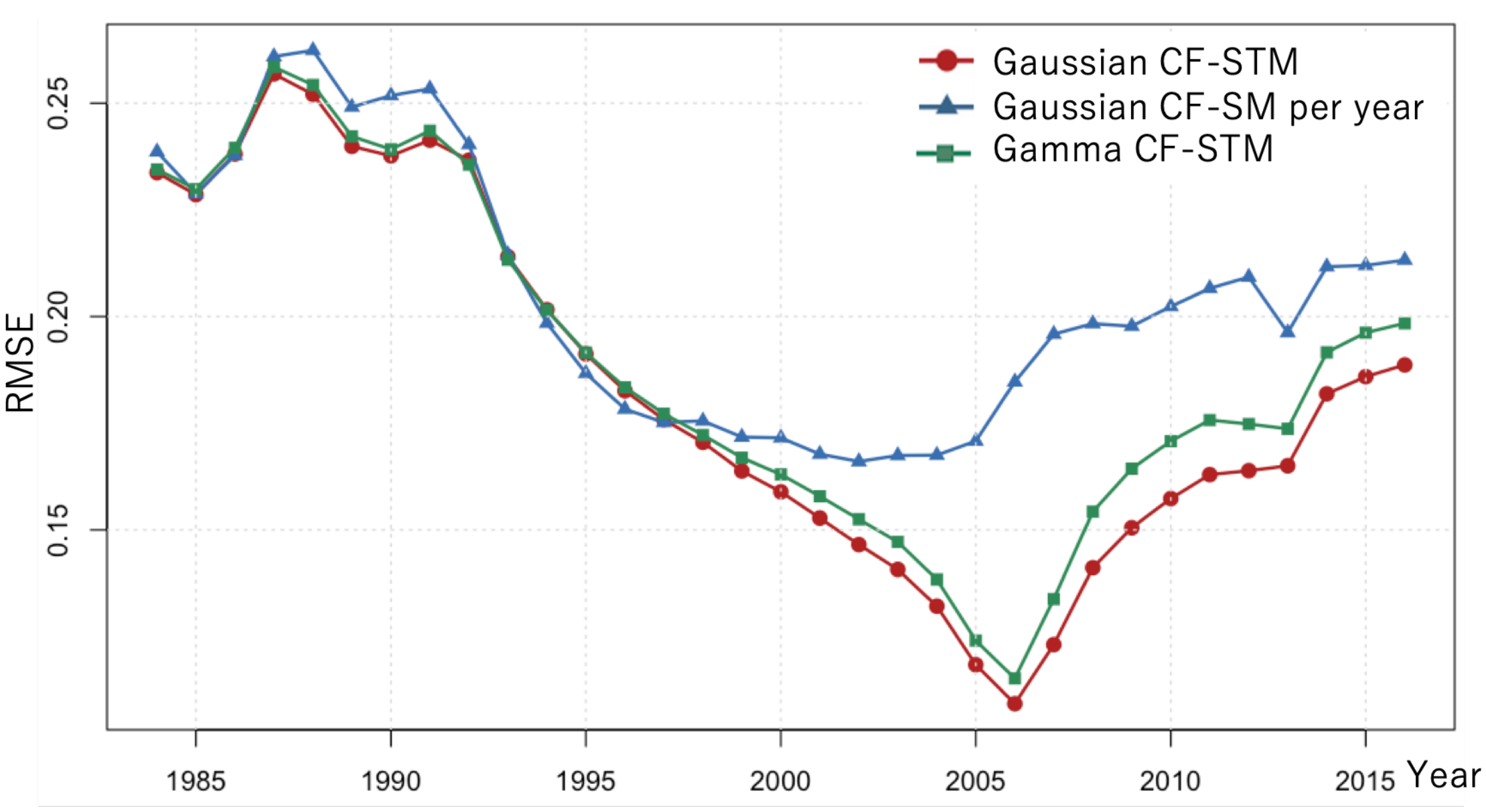

Figure 9: Comparison of yearly RMSE values.

## 5.5. Multiscale analysis

The multiscale processes shown in Figure 3 are useful for analyzing the multiscale structure underlying land prices and their temporal dynamics. For easier interpretation, the 55 scale-wise processes are aggregated into large-, moderate-, and local-scale components by aggregating the processes with bandwidth values of $h_r \geq 10$, $10 > h_r \geq 3$, and $3 > h_r$, respectively.

Large-scale processes substantially raised land prices across the study area in 1991, whereas moderate- and local-scale processes showed only minor temporal variations across years.

This suggests that the bubble economy, which peaked around 1991, broadly increased land prices across the region, with little increase at the moderate or local scales.

Throughout the study period, the large-scale process consistently indicated higher land prices in the southwestern region, including in the Shonan area (Figure 2), where coastal amenities and accessibility to Tokyo drove strong residential demand. The moderate-scale process highlighted higher prices along the coastline in the Shonan and Boso regions, reflecting the premiums associated with bayside areas. The moderate-scale process also increased prices around Tokyo, suggesting that Tokyo had a broader spatial influence than that of the other cities. Another city that strongly influenced moderate-scale processes is Tsukuba, a science city that developed around 1980 following the relocation of national research institutes. The moderate-scale process gradually increased around Tsukuba over time, likely reflecting its growing popularity and the opening of the Tsukuba Express Line in 2005.

The local-scale process increased land prices in central Tokyo as well as in selected suburban areas. These suburban hotspots tend to be located away from railway corridors, as shown in the bottom-right panel of Figure 10. This suggests that land prices in suburban areas located away from train stations are not as low as expected from the covariates. This may reflect that automobiles rather than railways are the principal modes of transportation in these areas.

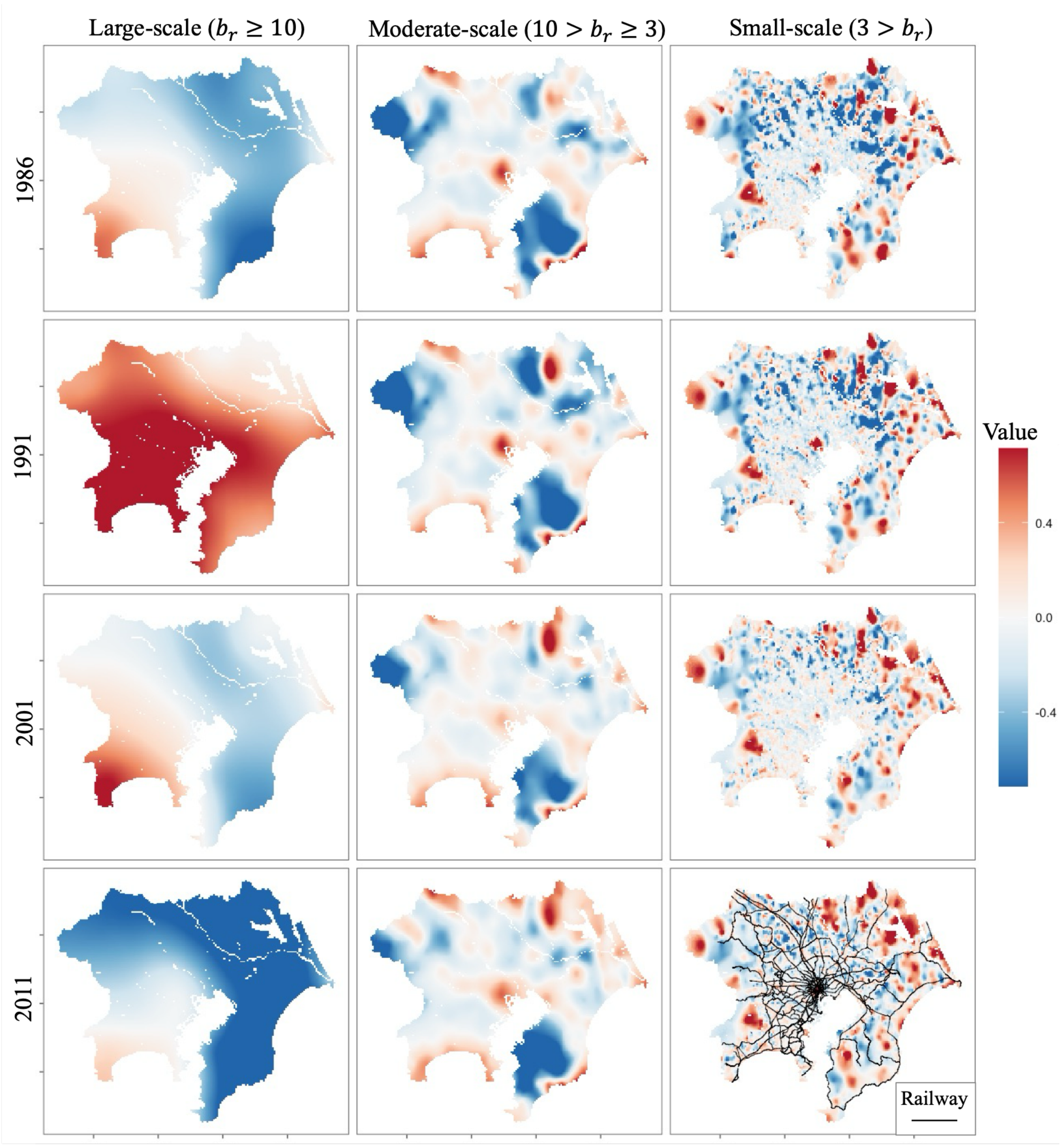

Figure 10: Large, moderate, and local-scale processes (1986, 1991, 2001, 2011). For visibility, the railway network is shown only in the bottom right panel.

## 6. Concluding remarks

This study developed coarse-to-fine spatiotemporal modeling (CF-STM), a scalable framework that extends the covariance-free spatial representation of CF-SM to spatiotemporal settings. By combining local spatial modeling with local state-space models, CF-STM avoids large covariance matrix operations while retaining the ability to represent complex latent spatiotemporal processes. The Monte Carlo experiments showed a predictive performance comparable to state-of-the-art SPDE-based models at a substantially lower computational cost, and the land price application demonstrated that CF-STM can flexibly recover dynamic spatiotemporal patterns underlying residential land prices.

Beyond its predictive accuracy and computational efficiency, an important contribution of CF-STM lies in its modular structure. Because multiscale spatial representation is separate from temporal inference, the framework is not intrinsically tied to a specific temporal modeling strategy. While this study focused on the linear-Gaussian case using the Kalman filter, the proposed formulation potentially provides a basis for future extensions incorporating more flexible sequential inference methods, including the ensemble Kalman filter (Evensen, 2003) and particle filter (Djuric et al., 2003), to handle nonlinear or non-Gaussian temporal dynamics. This space-time separation allows future developments in temporal inference and spatial modeling to proceed independently within a common framework. The cost of this modularity is that space-time dependence is not modeled jointly. Consequently, CF-STM does not guarantee the accuracy attainable by fully joint covariance-based models, as reflected in the slightly larger coefficient bias relative to SPDE in the binomial case.

Several extensions have naturally followed this framework. For example, local non-stationarity, including anisotropy (e.g., Wiens et al., 2020), can be incorporated by extending the specifications of each local model. Local models can also be generalized to locally weighted regressions, enabling spatially and temporally varying coefficients (e.g., Fotheringham et al., 2015) to be estimated within the CF-STM framework. In addition, the model aggregation function can be extended using alternative approaches, such as a mixture of experts and neural networks (e.g., Yang et al., 2023). Predictive variance, which currently assumes independence across scales and applies a global calibration, can be improved by introducing a scale-specific or spatially adaptive uncertainty calibration. Addressing these challenges would further strengthen CF-STM as a general, scalable, and interpretable framework for modeling large-scale spatiotemporal phenomena.

Code availability

CF-STM is implemented in an R package spCF (https://cran.r-project.org/web/packages/spCF/). Code and per-replicate results reproducing all tables in Section 4 and the appendices are available at https://github.com/dmuraka/spCF-cf_dglm-repro.

Appendix 1. Sensitivity of the SPDE benchmark to the mesh resolution

To examine how the accuracy and cost of the SPDE depend on the spatial mesh, we varied the mesh cutoff (the minimum allowed node spacing) on the moving-site DGP used for Table 2 (Gaussian response, 400 fresh sample sites per time point, $T = 40$, temporal dependence $\rho = 0.7$, spatial range 0.10, on the unit-square domain). For each cutoff, we evaluated the out-of-sample RMSE and CRPS, and the wall-clock fitting time per replicate, averaged over five replications.

Table A1 summarizes the evaluation results. Predictive accuracy saturates once the mesh resolves the spatial range (cutoff $\leq$ 0.10): the RMSE and CRPS are already 0.531 and 0.301 at cutoff = 0.10 and do not improve for finer meshes, matching CF-STM exactly. The fitting time, by contrast, grows roughly ten-fold each time the cutoff is halved, reaching 8,790.5 seconds per replicate at cutoff = 0.025. CF-STM attains the same accuracy in about 8.9 seconds, whereas at a comparable speed (cutoff = 0.20), SPDE is less accurate (RMSE 0.556). A still finer cutoff of 0.01 was not completed because a single fit was projected to exceed ten hours. These results support the cutoff = 0.05 adopted in the main text and reinforce that CF-STM matches the SPDE benchmark's predictive accuracy at a small fraction of its computational cost.

Table A1: Effect of the SPDE mesh cutoff on predictive accuracy (RMSE, CRPS), mesh size, and computation time on the moving-site Gaussian setting (five replications). CF-STM is shown as a mesh-free reference.

| Method | cutoff | RMSE | CRPS | Time (s) |
|---|---|---|---|---|
| CF-STM | – | 0.531 | 0.301 | 8.9 |
| SPDE | 0.200 | 0.556 | 0.312 | 8.5 |
| | 0.100 | 0.531 | 0.301 | 49.7 |
| | 0.050 | 0.530 | 0.301 | 503.2 |
| | 0.025 | 0.530 | 0.301 | 8790.5 |

Appendix 2. Monte Carlo experiments assuming Gaussian process-based DGP

The spatiotemporal process in the Monte Carlo experiments of Section 4 is built from a spatial moving average (Eq. 18), which does not exactly match the Matern covariance assumed by SPDE. To confirm that the conclusions do not favor CF-STM because of this choice, we repeated the experiments with a Gaussian-process-based DGP that is consistent with the SPDE model: at each time point the innovation field was drawn from a zero-mean Gaussian process with a Matern covariance (smoothness 1 and marginal variance 1), and the field evolved as a temporal AR(1) process; the two covariates were generated from the same Gaussian process. All other settings were unchanged. Tables A2 and A3 report the results.

Even under this GP-based DGP, which is exactly the covariance structure the SPDE benchmark assumes, CF-STM and SPDE remained essentially tied in predictive accuracy: their RMSE values were nearly identical, and CF-STM was marginally more accurate in the Gaussian case, whereas SPDE retained a small advantage in CRPS. Both clearly outperformed GLM, GAM, GSSM(-MS). For coefficient estimation, SPDE attained the smallest bias and standard deviation, with CF-STM close behind and clearly better than GLM and GAM. These findings coincide with those obtained under the spatial-moving-average process (Tables 1 and 3), indicating that the conclusions are robust to the choice of DGP.

Table A2: Predictive accuracy under the Gaussian-process DGP (bold: the first- and second-best methods).

| $\rho$ | Method | Gaussian | | Poisson | | Binomial | |
|---|---|---|---|---|---|---|---|
| | | RMSE | CRPS | RMSE | CRPS | RMSE | CRPS |
| 0.2 | GLM | 1.134 | 0.640 | 2.417 | 1.193 | 0.486 | 0.236 |
| | GAM | 1.036 | 0.584 | 2.253 | 1.114 | 0.478 | 0.229 |
| | GSSM | 0.680 | **0.388** | 1.931 | 0.933 | 0.469 | 0.220 |
| | GSSM-MS | 0.684 | 0.392 | 1.930 | 0.932 | 0.468 | 0.219 |
| | SPDE | **0.671** | **0.387** | **1.806** | **0.887** | **0.457** | **0.209** |
| | CF-STM | **0.666** | 0.404 | **1.810** | **0.891** | **0.458** | **0.210** |
| 0.7 | GLM | 1.477 | 0.834 | 2.405 | 1.188 | 0.486 | 0.236 |
| | GAM | 1.099 | 0.620 | 2.033 | 0.995 | 0.463 | 0.214 |
| | GSSM | 0.766 | **0.438** | 1.885 | 0.912 | 0.465 | 0.217 |
| | GSSM-MS | 0.771 | 0.443 | 1.890 | 0.913 | 0.465 | 0.217 |
| | SPDE | **0.755** | **0.435** | **1.787** | **0.875** | **0.453** | **0.205** |

| $\rho$ | Method | Gaussian | | Poisson | | Binomial | |
|---|---|---|---|---|---|---|---|
| | | RMSE | CRPS | RMSE | CRPS | RMSE | CRPS |
| | CF-STM | **0.735** | 0.451 | **1.791** | **0.880** | **0.454** | **0.206** |

Table A3: Estimation accuracy of the coefficient under the Gaussian-process DGP ($\rho = 0.7$).

| Method | Gaussian | | Poisson | | Binomial | |
|---|---|---|---|---|---|---|
| | Bias | Std.dev. | Bias | Std.dev. | Bias | Std.dev. |
| GLM | -0.001 | 0.046 | -0.002 | 0.025 | -0.109 | 0.035 |
| GAM | -0.004 | 0.031 | -0.003 | 0.017 | -0.068 | 0.029 |
| SPDE | 0.000 | 0.013 | -0.001 | 0.011 | -0.010 | 0.027 |
| CF-STM | 0.000 | 0.018 | -0.001 | 0.013 | -0.036 | 0.027 |

Appendix 3. The implied spatiotemporal covariance

We characterize the covariance that the multiscale field $\sum_{r=1}^{R} z_{t,r}(s)$ of CF-STM induces on the link scale, extending the purely spatial analysis of Murakami et al. (2026b) to the dynamic setting. Under an idealized process-convolution representation of the construction, the induced covariance is separable in space and time and, in a continuous-scale limit, reduces to a spatiotemporal Matérn covariance.

Throughout, we follow the notation of Section 3: $r = 1, \dots, R$ indexes the spatial scales, $h_r$ denotes the bandwidth at scale *r*, and $z_{t,r}(s)$ denotes the *r*-th scale process, with the sample index *i* suppressed because the continuum idealization below treats *s* as a generic location. We assume that (A1) each scale $r$ is generated by convolving in space a temporally autoregressive white-noise innovation $a_r(u,t)$, with $\mathrm{Cov}\big(a_r(u,t), a_r(u',t')\big) = \tau_r^2 \delta(u-u')\, \rho^{|t-t'|}$; (A2) the sampling density is approximately uniform, so normalized sums may be replaced by integrals; and (A3) the coarse-to-fine residual fitting renders distinct scales approximately uncorrelated. As in Section 3.2, a single pair ($\rho$, $v^2$) is shared across scales, so the AR(1) filters at all local centers share one temporal correlation. The spatial dimension is $p = 2$.

*A3.1. Per-scale covariance*

Writing scale $r$ in continuum form, $z_{t,r}(s) = \int k_{h_r}(\|s-u\|) a_r(u,t) \mathrm{d}u$ with kernel $k_h(\|u\|) = \exp\left(-\frac{\|u\|}{h}\right)$, which is the exponential kernel of Section 3.2, and substituting $u = h_r v$ as in the static case, gives exactly

$$Cov\left(z_{t,r}(s), z_{t\prime,r}(s')\right) = \sigma_r^2 \rho_0\left(\frac{d}{h_r}\right) \rho^{|\Delta t|}. \tag{A3.1}$$

where $d = \|s-s'\|$ , $\Delta t = t - t'$ , $\sigma_r^2 = \tau_r^2 h_r^2\, (k*k)(0)$ , and $\rho_0(\cdot) = \frac{(k*k)(\cdot)}{(k*k)(0)}$ is the normalized spatial kernel autoconvolution ($*$ denotes convolution). The temporal factor $\rho^{|\Delta t|} = \exp\left(-\frac{|\Delta t|}{\ell_t}\right)$, with $\ell_t = -\frac{1}{\log \rho}$, is the AR(1) correlation of the dynamic shared across local centers, a one-dimensional Matérn correlation of smoothness $\nu_t = 0.5$. In the stationary AR(1) case of Section 3.6, $\sigma_r^2$ corresponds to $\frac{v^2}{1-\rho^2}$. Each scale is therefore separable: the same spatial shape $\rho_0(\cdot)$ at the spatial scale $h_r$, multiplied by a common exponential temporal correlation. The bandwidth enters only through $\frac{d}{h_r}$, exactly as in the static case.

*A3.2. The separable spatiotemporal Matérn as a special case*

Summing the uncorrelated scales under (A3), and using that $\rho$ is shared across scales, the temporal factor separates:

$$C(d, \Delta t) = \sum_{r=1}^{R} \sigma_r^2 \rho_0\left(\frac{d}{h_r}\right) \rho^{|\Delta t|} = C_S(d) \cdot \exp\left(-\frac{|\Delta t|}{\ell_t}\right), \tag{A3.2}$$

where $C_S(d) = \sum_{r=1}^{R} \sigma_r^2 \rho_0\left(\frac{d}{h_r}\right)$ is the multiscale spatial covariance of CF-STM. The spatial marginal $C(d, 0) = C_S(d)$ is precisely the scale-mixture covariance of Murakami et al. (2026b): with Gaussian convolution kernels and Gamma-distributed scale weights $\sigma_r^2 \approx g\nu(\xi_r)$ evaluated at $\xi_r \propto h_r^2$, its continuous-scale limit is a spatial Matérn of smoothness $\nu_s$, through the Gaussian scale-mixture identity

$$(\alpha^2 + \|\omega\|^2)^{-(\nu+1)} = \frac{1}{\Gamma(\nu+1)} \int_0^{\infty} \xi^{\nu} \exp(-\alpha^2 \xi) \exp(-\xi \|\omega\|^2) d\xi. \tag{A3.3}$$

Combining the two factors,

$$C(d, \Delta t) \rightarrow \sigma^2 \cdot M_{\nu_s}(d\,;\, \ell_s) \cdot M_{0.5}(\Delta t\,;\, \ell_t) \tag{A3.4}$$

a separable spatiotemporal Matérn covariance whose spatial smoothness $\nu_s$ is set by the scale weights and whose temporal smoothness is fixed at 0.5 by the AR(1) dynamic. The spatial range is $\ell_s = \frac{1}{\lambda}$, where $\lambda$ is the rate of the Gamma scale weights in (A3.3), and the temporal range is $\ell_t = -\frac{1}{\log \rho}$. The finite set of scales selected by CF-STM gives a discrete approximation to this family.

Remark (positioning relative to the SPDE benchmark). Equation (A3.4) is the covariance family assumed by the separable SPDE spatiotemporal model (a spatial Matérn field evolving as a first-order autoregression), as implemented, for example, by sdmTMB with spatiotemporal = "ar1". CF-STM targets the same family through a finite multiscale approximation, which accounts for the close link-scale agreement between CF-STM and the SPDE benchmark reported in Section 4.

Remark (non-separability). Separability in (A3.2) follows from sharing a single $(\rho, v^2)$ across scales. Allowing a scale-dependent temporal correlation $\rho_r$ (equivalently $\ell_{t,r}$) instead yields $C(d, \Delta t) = \sum_{r=1}^{R} \sigma_r^2 \rho_0\left(\frac{d}{h_r}\right) \exp\left(-\frac{|\Delta t|}{\ell_{t,r}}\right)$, a mixture of separable covariances and hence a non-separable spatiotemporal model, in which coarse scales may persist longer in time than fine ones. This is a natural extension of the present construction.